\documentclass[preprint,notoc]{JHEP3}
\usepackage{epsfig}

\def\eslt{\not\!\!{E_T}}
\def\eslt{E_T^{\rm miss}}

\def\to{\rightarrow}

\def\bi{\begin{itemize}}
 \def\ei{\end{itemize}}
\def\te{\tilde e}
\def\c1p{C1^\prime}
\def\ta{\tilde a}
\def\tG{\tilde G}

\def\tu{\tilde u}

\def\ta{\tilde a}

\def\tb{\tilde b}

\def\tst{\tilde t}

\def\tg{\tilde g}

\def\tq{\tilde q}

\def\twpm{\tilde\chi^\pm}
\def\tz{\tilde\chi^0}
\def\alt{\stackrel{<}{\sim}}
\def\agt{\stackrel{>}{\sim}}
\def\be{\begin{equation}}  
\def\ee{\end{equation}}  
\def\bea{\begin{eqnarray}}  
\def\eea{\end{eqnarray}}  

\newcommand\njp[3]{{\it New\ J.\ Phys.\ }{\bf #1} (#2) #3}
\newcommand\annp[3]{{\it Annals\ Phys.\ }{\bf #1} (#2) #3}
\newcommand\sjp[3]{{\it Sov.\ J.\ Nucl.\ }{\bf #1} (#2) #3}

\def\Isajet{{\sc Isajet}}
\def\Softsusy{{\sc Softsusy}}

\title{Cosmological consequences of Yukawa-unified SUSY \\
with mixed axion/axino cold and warm dark matter}
\author{Howard Baer$^{a}$, Markus Haider$^{b,c}$, Sabine Kraml$^b$, 
Sezen Sekmen$^d$ and Heaya Summy$^{a}$\\
$^a$Dept.\ of Physics and Astronomy, University of Oklahoma, Norman, OK 73019, USA\\
$^b$Laboratoire de Physique Subatomique et de Cosmologie, UJF Grenoble 1, 
CNRS/IN2P3, INPG, 53 Avenue des Martyrs, F-38026 Grenoble, France\\
$^c$Inst.\ of Astro and Particle Physics, University of Innsbruck, A-6020 Innsbruck, Austria\\
$^d$Department of Physics, Middle East Technical Univ., TR-06531 Ankara, Turkey\\
E-mail: \email{baer@nhn.ou.edu}, \email{markus.haider@lpsc.in2p3.fr}, 
\email{sabine.kraml@lpsc.in2p3.fr}, \email{sezen.sekmen@cern.ch}, \email{heaya@nhn.ou.edu}}

\preprint{\vbox{LPSC 08-190}}

\abstract{
Supersymmetric models with $t-b-\tau$ Yukawa unification at $M_{GUT}$
qualitatively predict a sparticle mass spectrum including first and second generation scalars
at the 3--15 TeV scale, third generation scalars at the (few) TeV scale and gluinos 
in the sub-TeV range. 
The neutralino relic density in these models typically turns out to lie far above
the measured dark matter abundance, 
prompting the suggestion that instead dark matter is composed of an axion/axino mixture.
We explore the axion and thermal and non-thermal axino dark matter abundance in 
Yukawa-unified SUSY models. We find in this scenario that 
{\it i}).\ rather large values of Peccei-Quinn symmetry breaking scale 
$f_a\sim 10^{12}$ GeV are favored and {\it ii}).\ rather large values of 
GUT scale scalar masses $\sim 10-15$ TeV allow for the re-heat temperature 
$T_R$ of the universe to be $T_R\agt 10^6$ GeV.
This allows in turn a solution to the gravitino/Big Bang Nucleosynthesis problem 
while also allowing for baryogenesis via non-thermal leptogenesis. 
The large scalar masses for Yukawa-unified models are also favored by 
data on $b\to s\gamma$ and $B_s\to \mu^+\mu^-$ decay.
Testable consequences from this scenario include a variety of robust LHC signatures,
a possible axion detection at axion search experiments, but null results from
direct and indirect WIMP search experiments.
}

\begin{document}

\section{Introduction}
\label{sec:intro}

The celebrated unification of gauge couplings at scale $M_{GUT}\simeq 2\times 10^{16}$ GeV
under minimal supersymmetric standard model (MSSM) renormalization group evolution\cite{gaugeunif}
strongly suggests that nature is described by some sort of supersymmetric grand unified theory (SUSY GUT) 
model at very high energy scales.
While the GUT gauge group $SU(5)$\cite{su5} has many compelling and beautiful properties, the gauge group
$SO(10)$ has an even greater appeal\cite{so10}, 
and also some indirect experimental support in terms of how well
see-saw neutrino mass fits into the general scheme\cite{seesaw}. 
While both $SU(5)$ and $SO(10)$ SUSY GUT theories admit gauge coupling
unification, $SO(10)$ theories also yield {\it matter unification}, in that all
superfields of a single Standard Model (SM) generation-- plus a SM gauge singlet superfield $\hat{N}^c$
containing a right-hand neutrino state-- fit neatly into the 16-dimensional spinorial
representation $\hat{\psi}_{16}$ of $SO(10)$. 

In the simplest $SO(10)$ SUSY GUT models, the MSSM Higgs
superfields-- $\hat{H}_u$ and $\hat{H}_d$-- both live in the fundamental representation $\hat{\phi}_{10}$.
In these models, the superpotential has the form
\be
\hat{f}\ni f\hat{\psi}_{16}\hat{\psi}_{16}\hat{\phi}_{10} +\cdots
\ee 
where the dots represent model-dependent terms which, for instance, might include further Higgs fields
responsible for the GUT gauge symmetry breaking. The coupling $f$ represents the {\it unified} Yukawa coupling
of each generation: thus, just as $SU(5)$ models often predict $f_b-f_\tau$ unification, simple
$SO(10)$ SUSY GUT models predict the more restrictive $t-b-\tau$ 
Yukawa coupling unification at the GUT scale\cite{old}.

It is probably fair to say that at this moment no compelling $SO(10)$ SUSY GUT model yet exists.
Models based in four spacetime dimensions require large, unwieldy Higgs representations to break the 
$SO(10)$ GUT symmetry. Newer models formulated in extra spacetime dimensions are able to do away with the 
large Higgs reps and break the GUT symmetry via compactification of the extra dimensions\cite{exdimguts}.
In our approach here, we will adopt a pragmatic view, assuming that the MSSM (or MSSM plus gauge singlets)
is the correct effective theory describing physics between the weak and GUT scales, and we will explore
some of the consequences of requiring the three third generation Yukawa couplings to have a high degree
of unification at $M_{GUT}$, as suggested by simple $SO(10)$ SUSY GUT models.

In our calculation to check third generation Yukawa coupling unification,
we begin
with the measured gauge couplings and third generation fermion masses, and use renormalization group
methods to evolve the coupled gauge and Yukawa couplings up to the GUT scale. The calculation
ends up being sensitive to the entire SUSY particle mass spectrum through weak scale threshold corrections
involved in transitioning between the SM and MSSM effective field theories\cite{hrs}.
The parameter space of the model is given by
\be
m_{16},\ m_{10},\ M_D^2,\ m_{1/2},\ A_0,\ \tan\beta, \ {\rm sign}(\mu )
\label{eq:pspace}
\ee
where $m_{16}$ is the GUT scale mass of all matter scalars, $m_{10}$ is the GUT 
scale mass of Higgs scalars, $M_D$ is their D-term value, 
$m_{1/2}$ is the (unified) GUT scale gaugino mass, $A_0$ is the unified GUT scale SSB trilinear term,
$\tan\beta\equiv v_u/v_d$ is the weak scale ratio of Higgs field vevs, and $\mu$ is the superpotential
Higgs bilinear term, whose magnitude--but not sign--is determined by the scalar potential
minimization conditions. 

In practice, the two Higgs field soft breaking terms-- $m_{H_u}^2$ and 
$m_{H_d}^2$-- cannot be degenerate at $M_{GUT}$ and still allow for an appropriate radiative breakdown of 
electroweak symmetry (REWSB). Effectively, $m_{H_u}^2$ must be less than $m_{H_d}^2$ at $M_{GUT}$ in order
to give $m_{H_u}^2$ a head start in running towards negative values at $M_{weak}$. We parametrize 
the Higgs splitting as $m_{H_{u,d}}^2=m_{10}^2\mp 2M_D^2$ in accord with nomenclature for
$D$-term splitting to scalar masses when a gauge symmetry undergoes a breaking which reduces the 
rank of the gauge group. A $D$-term splitting should apply to matter scalar SSB terms as well; in practice,
better Yukawa unification is found when the splitting is only applied to the Higgs SSB terms. 
Such a GUT scale Higgs mass splitting might arise via GUT scale threshold corrections\cite{bdr2}.

In previous work, the above parameter space was scanned over (via random scans\cite{bf,abbbft} and also
by more efficient Markov Chain Monte Carlo (MCMC) scans\cite{bkss}) to search for Yukawa unified solutions
using the Isasugra subprogram of \Isajet\cite{isajet} for sparticle mass computations.
The quantity
\be
R=\frac{max(f_t,f_b,f_\tau )}{min(f_t,f_b,f_\tau )}\ \ \ ({\rm evaluated\ at\ Q=M_{GUT}}),
\label{eq:R}
\ee
was examined, where solutions with $R\simeq 1$ gave apparent Yukawa coupling unification.
For superpotential Higgs mass parameter $\mu >0$ (as favored by $(g-2)_\mu$ measurements),
Yukawa unified solutions with $R\sim 1$ were found
but only for {\it special} choices of GUT scale boundary conditions\cite{bf,bdr1,bdr2,abbbft,drrr,bkss}:
%
\be
  m_{16}\sim 3-15~{\rm TeV},~
  A_0\sim -2m_{16}\,,~
  m_{10}\sim 1.2 m_{16}\,,~
  m_{1/2}\ll m_{16}\,,~
  \tan\beta \sim 50\,.
  \label{eq:HSpar}
\ee
Models with this sort of boundary conditions were derived even earlier in the context
of inverted scalar mass hierarchy models (IMH) which attempt to reconcile suppression of
flavor-changing and $CP$-violating processes with naturalness 
via multi-TeV first/second generation and sub-TeV scale third generation scalars\cite{bfpz}.
The Yukawa-unified spectral solutions were thus found in Refs.~\cite{abbbft,bkss} to occur with the 
above peculiar choice of boundary conditions as long as $m_{16}$ was in the multi-TeV regime.

Based on the above work\cite{abbbft,bkss}, the sparticle mass spectra from Yukawa-unified SUSY 
models are characterized qualitatively by the following conditions:
\begin{itemize}
\item first and second generation scalars have masses in the $3-15$ TeV regime,
\item third generation scalars, $\mu$ and $m_A$ have masses 
in the TeV to few TeV regime 
(owing to the inverted scalar mass hierarchy),
\item gauginos, including the gluino, have sub-TeV masses,
\item the lightest neutralino $\tz_1$ is nearly pure bino with mass typically $m_{\tz_1}\sim 50-80$ GeV.
\end{itemize}
The presence of a bino-like $\tz_1$ along with multi-TeV scalars gives rise to a neutralino
cold dark matter (CDM) relic abundance that is typically in the range
$\Omega_{\tz_1}h^2\sim 10-10^4$, {\it i.e.} 
far above\cite{auto} the WMAP measured\cite{wmap5} value 
\be
\Omega_{CDM}h^2=0.110\pm 0.006\ \ \ \ \  
\ee
by several orders of magnitude.

One solution to the CDM problem in Yukawa-unified SUSY models 
occurs for $m_{16}\sim 3$~TeV, where one can have $m_{\tz_1}\simeq m_h/2$.
In this case, the $\tz_1$ would pair-annihilate in the early universe 
through the light Higgs $h$ resonance at a sufficient rate to obtain the desired relic 
density\cite{bkss}, and could hence be the stable lightest SUSY particle (LSP). 
However, since $m_{16}$ is relatively low, Yukawa coupling unification only occurs 
at the $R\sim 1.09$ level.\footnote{Another possibility, having $m_A$ light enough that
$\tz_1\tz_1$ can annihilate through the $A$ resonance, appears to be excluded because these
cases violate limits on $BF(B_s\to\mu^+\mu^- )$ decay\cite{bkss}.} 

Another very compelling way out of the Yukawa-unified dark matter abundance problem occurs
if one makes the additional assumption that the strong $CP$ problem is 
solved by the Peccei-Quinn mechanism\cite{pq}, which leads to the presence 
of a light pseudoscalar particle, the axion~$a$~\cite{ww,axreview}. 
Since we are working in a supersymmetric theory, the axion occurs as part of an axion 
supermultiplet\cite{nillesraby}, 
which contains not only the axion, but a spin-0 saxion (with mass of order the weak scale), and a
spin-$1\over 2$ {\it axino} $\ta$. The axino is $R$-parity odd. While the saxion is 
expected to have a mass of order the SUSY breaking scale, the axino mass is very model dependent,  and can lie
anywhere in the keV-GeV range\cite{axmass,ckkr}. 
The axino then can serve as the  LSP instead of the lightest neutralino\cite{wilczek,steff_rev}.

In the case of an axino LSP, the supposed neutralino relic
abundance is greatly reduced since $\tz_1\to \ta\gamma$ decay can occur with a 
lifetime in the range of $\sim 10^{-5}-10^{1}$ sec (depending on parameters). 
This decay time is sufficiently short that late-time neutralino decay to axino in the early universe
should not upset successful predictions of Big Bang Nucleosynthesis (BBN)\cite{ckkr}.
The neutralino abundance then gets reduced by the 
ratio $m_{\ta}/m_{\tz_1}$ which can be in the range $10^{-1}-10^{-4}$. 
The $\ta$ coming from $\tz_1$ decay would actually
constitute {\it warm} dark matter as long as $m_{\ta}\alt 1$ GeV\cite{jlm}. 
However, axinos can also be produced thermally in the early universe, and these would
consititute CDM as long as $m_{\ta}\agt 100$ keV\cite{ckkr,steffen}. 
Thus, in the axino LSP case, we could have a mixed dark matter (DM) scenario\cite{julien} with 
\bi
\item thermally produced cold or warm axino DM, 
\item an admixture of warm axino DM arising from $\tz_1\to\ta\gamma$ decays and 
\item a possibly large presence of cold {\it axion} DM.
\ei

The axino LSP scenario turns out to be even more compelling cosmologically than just as a means 
to reconcile the dark matter relic abundance with Yukawa-unified models.
In this class of Yukawa-unified solutions with $m_{16}$ in the multi-TeV range, we expect from 
supergravity theory that scalar SSB terms should be directly related to the gravitino mass $m_{3/2}$, 
and so we also expect the gravitino $\tG$ to lie in the multi-TeV range. 
The cosmological gravitino problem-- wherein gravitinos produced
thermally in the early universe suffer a late-time decay, thus destroying the successful predictions
of BBN-- can be avoided. For $m_{3/2}\alt 5$ TeV, the re-heat temperature $T_R$
must be $T_R\alt 10^5$ GeV\cite{kohri}, thus creating tension with most viable 
mechanisms for baryogenesis\cite{buchm}. 
However, for
$m_{3/2}\agt 5$ TeV, the re-heat bound is much higher: $T_R\alt 10^8-10^9$ GeV. 
This range of $T_R$ is too low for thermal leptogenesis (which requires $T_R\agt 10^{10}$ GeV)\cite{buchm}, 
but is exactly what is needed for baryogenesis via {\it non-thermal} leptogenesis\cite{ntlepto},
wherein the heavy right-hand neutrino states are not produced thermally, but rather via inflaton decay.
It was pointed out in Ref. \cite{bs} that this is also the exact range needed to generate a dominantly {\it cold} 
axino DM universe. Thus, the
whole scenario fits together to offer a consistent cosmological picture of BBN, non-thermal leptogenesis 
and CDM composed of axions and/or axinos, and one solves the strong $CP$ problem to boot\cite{bs}!

In this paper, we make a detailed study of the cosmological consequences of Yukawa-unified SUSY GUT models
with an axino LSP. 
The initial study proposed in Ref.~\cite{bs} assumed only a negligible
component of axion dark matter.
In this study, we now fold in the axion contribution to the dark matter abundance. 
In addition, we have updated the value of $m_t$ in our calculations, and refined the Yukawa coupling 
1-loop beta function threshold effects in \Isajet . We also incorporate in this study the results of our 
Markov Chain Monte Carlo (MCMC) approach to finding Yukawa-unified solutions. Whereas the 
Yukawa-unified solutions found via a random scan in Ref.~\cite{bs} had $m_{16}\sim 15-20$ TeV, 
the more efficient MCMC scans used here are able to find many solutions for $m_{16}$ values
as low as $3-10$~TeV.

The paper is organized as follows.
In Sec.~\ref{sec:mass}, we update our sparticle mass predictions for Yukawa-unified SUSY
models from \Isajet\ using 1. an improved beta-function threshold decoupling, 2. an updated 
value of the top mass $m_t=172.6$ GeV, and 3. Markov Chain Monte Carlo (MCMC) scans of the parameter space. 
We also exhibit plots of the $b\to s\gamma$ and $B_s\to\mu^+\mu^-$ branching fractions versus
$m_{16}$ and find these favor $m_{16}$ values $\agt 10$ TeV.
Moreover, we perform 
MCMC scans for Yukawa-unified solutions with an alternative spectrum generator, \Softsusy, 
and compare these results with those gained from \Isajet.
In Sec.~\ref{sec:dm}, we review elements of axion and axino dark matter cosmology, 
including plots of the axion relic abundance, axino relic abundance and neutralino lifetime.
In Sec.~\ref{sec:axdm}, we discuss the gravitino/BBN problem, non-thermal leptogenesis 
via inflaton decay, and mixed axion/axino dark matter.
Our main findings are located in Sec.~\ref{ssec:results}. Here, we calculate all three components of
axion/axino dark matter in Yukawa-unified models for $m_{16}$ values of 5, 8, 10 and 15 TeV.
We explore scenarios wherein axions constitute either most, or hardly any, of the dark matter abundance.
Our results are presented in the $m_{\ta}$ vs. $T_R$ plane. We find that models with a 
rather large value of the Peccei-Quinn breaking scale $f_a\sim 10^{12}$ GeV are favored, as
well as models with $m_{16}\sim m_{3/2}$ on the high side: $\sim 10-15$ TeV.
In these cases, the re-heat temperature of the universe can range above
$10^6$ GeV, allowing for a solution to the BBN gravitino problem as well as allowing for
non-thermal leptogenesis.
Our conclusions are presented in Sec.~\ref{sec:conclude}.
In an Appendix, we list updated Yukawa-unified benchmark points from 
\Isajet\ 7.79.

\section{Updated spectrum predictions for Yukawa-unified SUSY models}
\label{sec:mass}

\subsection{Updated Isajet calculation of sparticle mass spectra}
\label{ssec:isajet}

The \Isajet\ calculations begin by adopting the fermion mass boundary conditions that 
$m_b^{\overline{DR}}(M_Z)=2.83$ GeV, $m_\tau^{\overline{DR}}(M_Z)=1.7463$ GeV, and 
$m_t(pole)=172.6$ GeV, 
along with the measured gauge couplings; in particular we 
take $\alpha_s^{\overline{MS}}(M_Z)=0.1172$.
Note the value of $m_t$ 
represents an update due to recent D0 and CDF measurements\cite{mtop} over our previous work 
Ref.~\cite{bkss} which was done using $m_t=171$ GeV. 
Our results do not change qualitatively upon varying the fermion masses within their error bars\cite{abbbft}.
We use the \Isajet\,7.79\cite{isajet} program to perform two loop
RG evolution of gauge and Yukawa couplings and all soft SUSY breaking terms. 
For gauge and Yukawa couplings, \Isajet\ actually uses an RGE approach wherein the 1-loop beta functions
change whenever a SSB threshold is passed over\cite{castano,box}. The Yukawa coupling evolution depends on
finite terms from 1-loop MSSM threshold effects\cite{hrs,bmpz}; these are implemented at a scale
$Q=M_{SUSY}\equiv\sqrt{m_{\tst_L}m_{\tst_R}}$. 
The threshold effects cause the entire calculation to depend sensitively on the
sparticle mass spectrum, which enters the various loop corrections to $f_t$, $f_b$ and $f_\tau$.
In addition, in the two-loop RG running\cite{mv} of SSB terms from $M_{GUT}$ to the weak scale, 
non-mixing soft terms are frozen out at their own mass scale, 
while SSB terms that mix are frozen at the scale $M_{SUSY}$\cite{bfkp}.
Complete 1-loop corrections are then applied to all sparticle masses. This approach leads to good
agreement with other publicly available sparticle spectra codes\cite{kraml} when the sparticle masses
are all nearby in mass scale. For spectra suffering severe splitting (as will be the case here),
the \Isajet\ multiple decoupling approach attempts to deal with the fact that several mass
scales may be present in and around the weak scale\cite{box}. 

Two updates in the \Isajet\,7.79 code affect the running of Yukawa couplings.
In early versions of \Isajet, all squark contributions to RG running were decoupled at a 
common squark mass scale taken to be $m_{\tu_L}$ and all sleptons were decoupled at a common
scale taken to be $m_{\te_L}$. In \Isajet\,7.79, the first/second and third generation
squarks and sleptons decouple at the values of the corresponding soft SUSY breaking terms:
thus, in a case where $m_{16}=10$ TeV, first/second generation squarks and sleptons decouple 
around $10$ TeV while third generation squarks and sleptons decouple at a much lower scale
around 3 TeV.
In addition, earlier versions of \Isajet\ included 2-loop terms for MSSM running between 
$M_{SUSY}$ and $M_{GUT}$, but turned these off
for $Q<M_{SUSY}$, where the SM was the expected effective theory. In \Isajet\,7.79, the MSSM 
two-loop terms remain in for $Q<M_{SUSY}$, since the scale of decoupling of 2-loop terms is a 3-loop effect.
The current version should give a better estimate of Yukawa coupling evolution in models
with severe first/second and third generation splitting, as in Yukawa-unified models.

\subsection{Isajet/Softsusy comparison for Yukawa-unified spectra}
\label{ssec:isasoft}

Other public spectrum codes follow a different approach than \Isajet\ and perform a 1-step 
decoupling of SUSY particles at $Q=M_{SUSY}$. This ``all-at-once'' transition may lead to 
some differences in the sparticle masses in particular in the case of a widely split spectrum\cite{bfkp}. 
It is therefore interesting to compare results from different spectrum codes. 
Here, we choose \Softsusy\,2.0.18\cite{softsusy} for a representative comparison 
of the 1-step decoupling with the multi-step approach implemented in \Isajet\,7.79.

\FIGURE[t]{
\includegraphics[width=9cm]{yuk_comp}
\caption{Evolution of $f_t$, $f_b$ and $f_\tau$ from the weak scale to 
the GUT scale for the point in Table~\ref{tab:compare}. 
\Isajet\ results are solid, while \Softsusy\ results are dashed.
The large jumps around 3 TeV from \Isajet\ 
correspond to the SM--MSSM threshold corrections.
}\label{fig:yuk779}}

The evolution of Yukawa couplings between the electroweak and GUT scales
is illustrated in Fig.~\ref{fig:yuk779} for a point with $m_{16}=8$~TeV, 
which runs for both \Isajet\ and \Softsusy.
As can be seen,  
\Softsusy\ implements threshold corrections to the [gauge and] Yukawa couplings 
at scale $Q=M_Z$, and then runs from $M_Z$ to $M_{GUT}$ with full MSSM RG evolution.
The SM--MSSM threshold corrections in \Isajet\ occur at $Q=M_{SUSY}$, and give 
rise to the the discontinuities located around $M_{SUSY}\sim 3$ TeV.
The steep slope for $f_t$ and $f_b$ for $Q<M_{SUSY}$ occurs mainly because the
co-efficient of the QCD $g_s^2$ contribution to $f_{t/b}$ running changes from
$8$ in the SM for $Q<M_{SUSY}$ to $16/3$ in the MSSM for $Q>M_{SUSY}$.
The two programs obviously differ in the bottom Yukawa coupling $f_b$.
Part of this difference comes from somewhat different weak scale boundary 
conditions on the value of $m_b^{\overline{DR}}(M_Z)$: as mentioned above, 
\Isajet\ has a hard-coded value of $m_b^{\overline{DR}}(M_Z)=2.83$~GeV;
in \Softsusy, on the other hand, $m_b^{\overline{DR}}(M_Z)$ is computed from 
$m_b^{\overline{MS}}(m_b)$ with SUSY corrections added at $M_Z$.\footnote{To 
comply with the bottom mass used by \Isajet\ in other parts of the program, 
we use $m_b^{\overline{MS}}(m_b)=4.2$~GeV as input in \Softsusy, leading to  
$m_b^{\overline{DR}}(M_Z)\simeq 2.6$~GeV depending on the exact parameter point.} 
The larger value of $f_b$ in \Softsusy\ causes $f_\tau$ to run to slightly higher values
at $M_{GUT}$ than for \Isajet\ .
The parameters of the point used and resulting mass spectra are listed 
in Table~\ref{tab:compare}.
\Softsusy\ gives Yukawa unification at 2\%, \Isajet\ gives Yukawa unification at 8\%.
While many masses are very similar, the values of $m_{\tg}$ and especially $m_A$
differ quite a bit.


%
\begin{table}\centering
\begin{tabular}{lcc}
\hline
parameter & \Softsusy\ & \Isajet\ \\ 
\hline
$f_t$      & 0.559 & 0.555 \\ 
$f_b$      & 0.549 & 0.512 \\ 
$f_\tau$   & 0.560 & 0.526 \\ 
$R$        & 1.02  & 1.08  \\ 
$\mu$      & 2130 & 2178  \\
$m_{\tg}$   & 332 & 383   \\
$m_{\tu_L}$ & 8004 & 7973   \\
$m_{\tst_1}$& 1892 & 1812   \\
$m_{\tb_1}$ & 2487 & 2653  \\
$m_{\te_R}$ & 8064 & 8065  \\
$m_{\twpm_1}$ & 116 & 121  \\
$m_{\tz_2}$ & 120 & 120  \\ 
$m_{\tz_1}$ & 53.4 &  53.1  \\ 
$m_A$       & 681 &  1955  \\
$m_h$       & 129.1 &  127.5  \\ \hline
\end{tabular}
\caption{Masses in~GeV units and parameters
for a Yukawa-unified point using \Softsusy\,2.0.18,
\Isajet\,7.79. 
We take $m_{16}=8000$ GeV, $m_{10}=9760.3$ GeV,
$M_D=2435.7$ GeV, $m_{1/2}=67.8752$ GeV, $A_0=-16007$ GeV, 
$\tan\beta =48.71$ and $m_t=172.6$ GeV.
Thus, $m_{H_d}=10350.3$ GeV and $m_{H_u}=9132.3$ GeV
}
\label{tab:compare}
\end{table}

\FIGURE[t]{
\includegraphics[width=10cm]{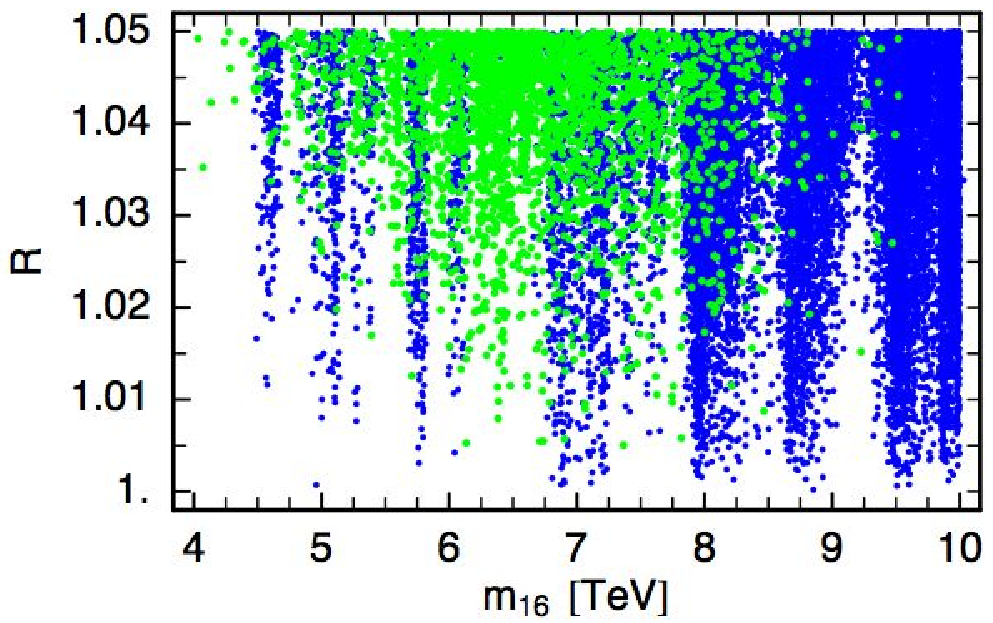}
\caption{Degree of Yukawa unification, $R$, versus $m_{16}$ from MCMC scans 
over the $SO(10)$ model parameter space. 
\Isajet\,7.79 results are in blue, while \Softsusy\,2.0.18 results are in green.
}\label{fig:pm16R}}

We next scan over the parameter space Eq.~(\ref{eq:pspace}) using a MCMC
algorithm, as detailed in \cite{bkss}, that searches for solutions with $R$ as low as possible.
Figure~\ref{fig:pm16R} shows the degree of Yukawa unification, $R$, found with the two codes 
as a function of $m_{16}$. Points from \Isajet\,7.79 are shown in blue, while points 
from \Softsusy\,2.0.18 are shown in green. As one can see, with both programs a high degree 
of Yukawa unification can be found, and these solutions tend to prefer high $m_{16}$.
However, as is apparent from the density of points, solutions with low $R$ are more 
easily found with \Isajet\ than with \Softsusy. Another important difference is that \Softsusy\ 
generates Yukawa-unified solutions only for $m_{16}$ up to about $9.5$~TeV, 
while \Isajet\ generates solutions for $m_{16}$ well beyond $10$~TeV 
(here we only show results up to 10 TeV).
The reason lies in the fine-tuning of the $\mu$ parameter and the electroweak symmetry breaking.

\FIGURE[t]{
\begin{tabular}{rr}
\includegraphics[width=7cm]{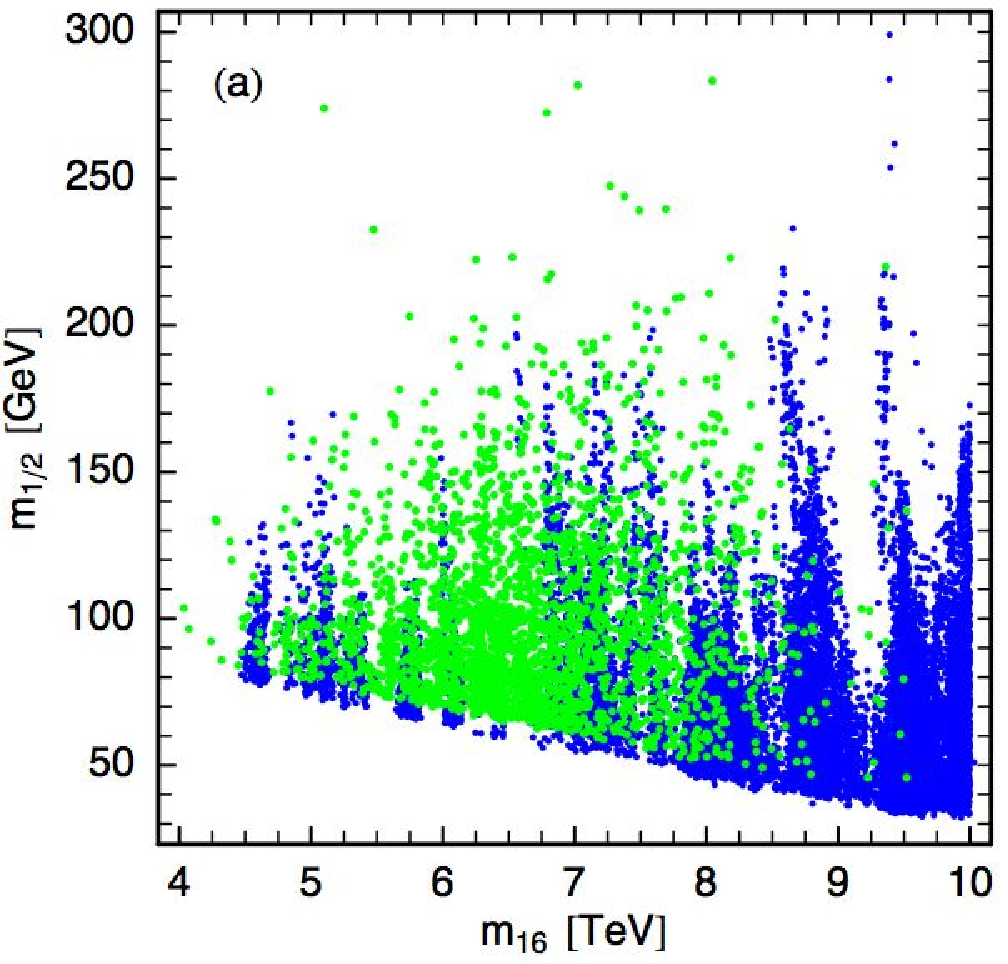} & 
\includegraphics[width=7.2cm]{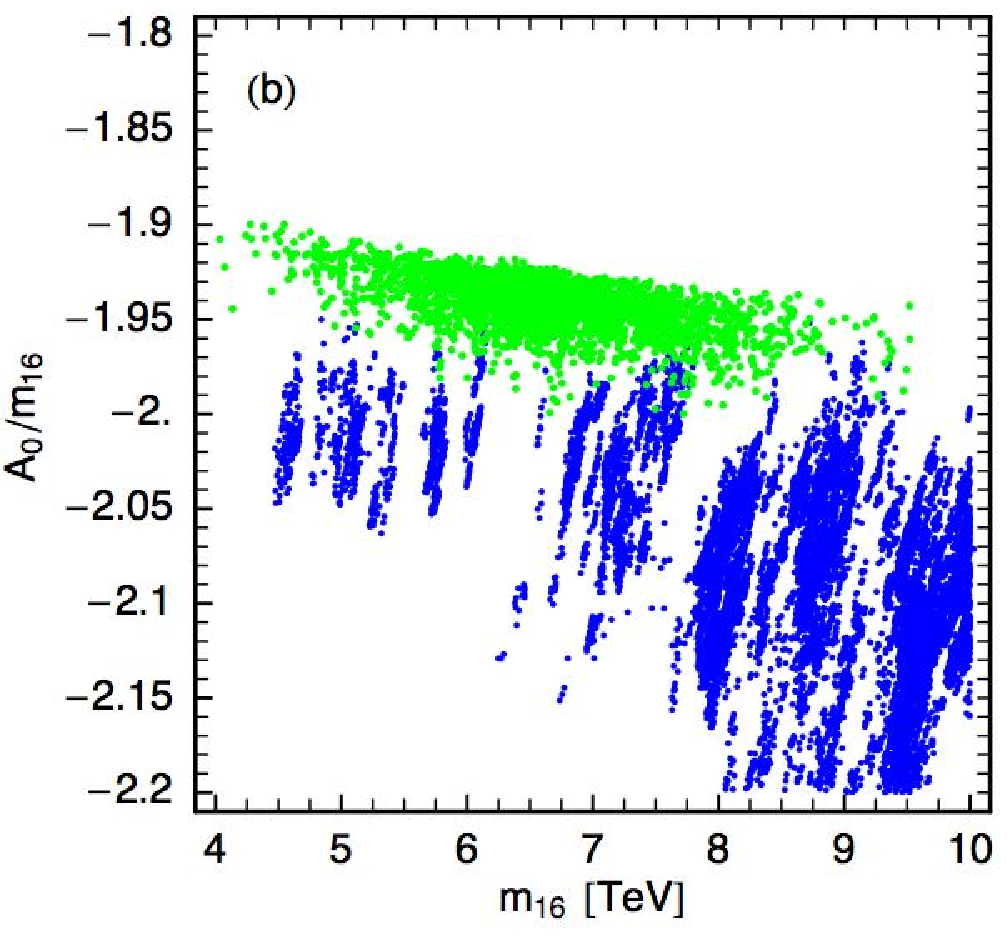} \\
\includegraphics[width=6.8cm]{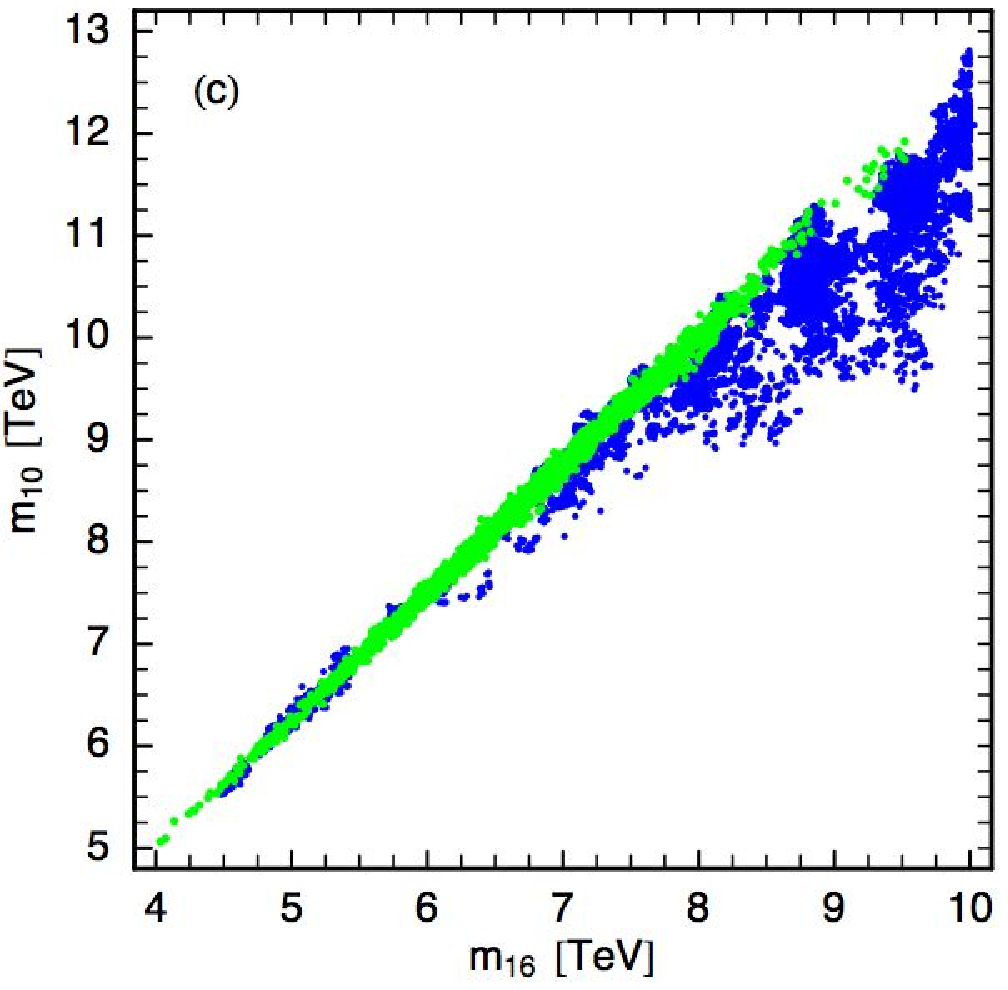} &
\includegraphics[width=6.65cm]{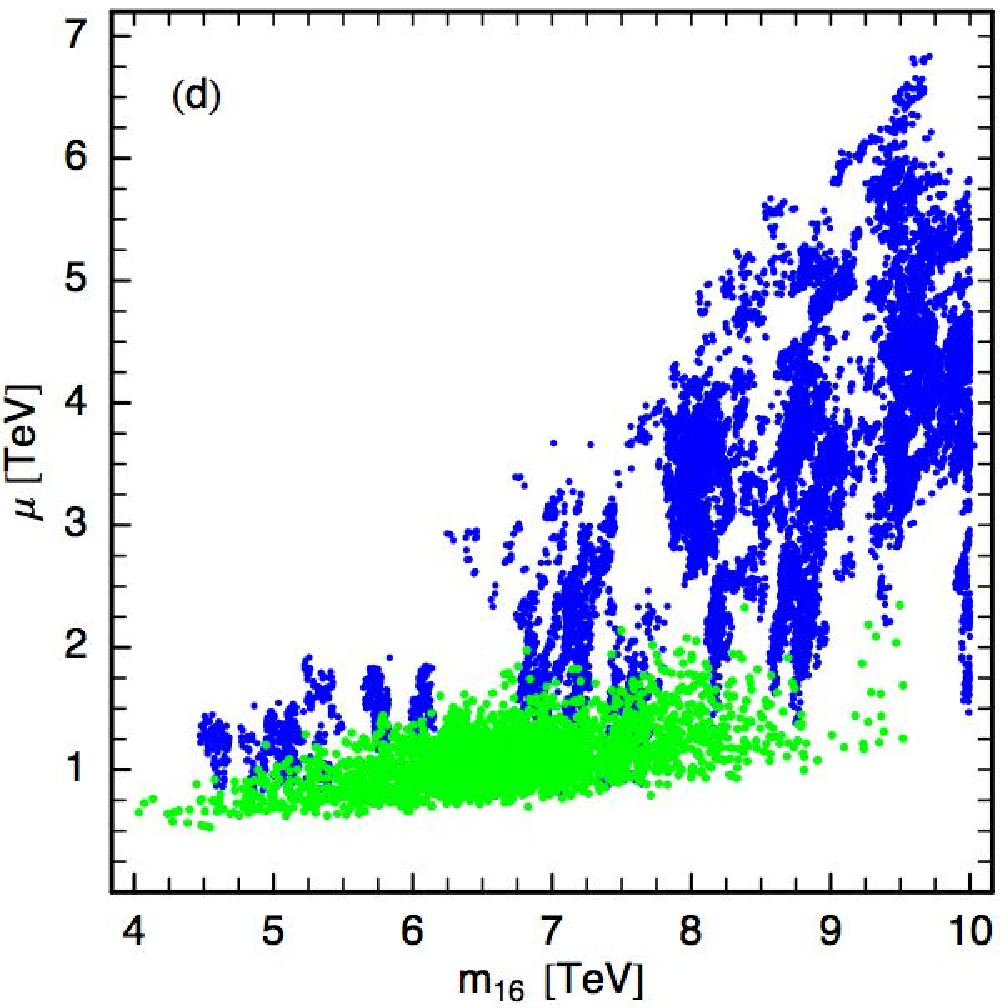}
\end{tabular}
\caption{Points with $R\le1.05$ from MCMC scans 
projected into various planes:  
(a) $m_{1/2}$ vs.\ $m_{16}$, (b)  $A_0/m_{16}$ vs.\ $m_{16}$,  (c) $m_{10}$ vs.\ $m_{16}$, 
and (d) $\mu$ vs.\ $m_{16}$. 
\Isajet\,7.79 results are in blue, while \Softsusy\,2.0.18 results are in green.
}\label{fig:parspace}}

Overall, Yukawa-unified solutions obtained with either program obey the same qualitative 
conditions Eq.~(\ref{eq:HSpar}). This is illustrated in detail in Figs.~\ref{fig:parspace}(a)--(d), 
which show MCMC scan points with $R\le1.05$ projected into various planes. 
In particular, frame (a) shows that $m_{16}$ is in the multi-TeV range while $m_{1/2}\sim 100$~GeV.
From frame (b), which shows the $A_0/m_{16}\ vs.\ m_{16}$ plane, we see that both programs
require $A_0\sim -2m_{16}$, although \Isajet\ prefers somewhat more negative values of 
$A_0$ than \Softsusy\ does.
Frame (c) illustrates the correlation $m_{10}\sim 1.2 m_{16}$. 
Finally, frame (d) shows an important difference in results from the two codes: 
in spectra obtained with \Isajet\ the $\mu$ parameter grows with $m_{16}$ and reaches 
values of a few TeV, 
while in spectra obtained with \Softsusy, the value of $\mu$ remains around 1--2 TeV.

In Fig.~\ref{fig:prelic}(a), we show the value of neutralino relic density from 
both \Isajet\ and \Softsusy\ versus $m_{16}$ for points with $R\le1.05$. 
The relic density is calculated with {\sc Micromegas}\cite{micromegas}, 
which easily interfaces with both \Isajet\ and \Softsusy.
Here, we see \Softsusy\  
predicts $\Omega_{\tz_1}h^2\sim 1-1000$, with some points extending down to
$\Omega_{\tz_1}h^2\sim 0.3$. These latter points occur due to neutralino
annihilation near the light Higgs $h$ resonance, but do not quite reach the
WMAP-measured dark matter density because the higgsino fraction $f_H$ of the $\tz_1$ 
is only few per mil. 
\Isajet\ tends to give an even smaller higgsino fraction 
(larger $\mu$ parameter, c.f. Fig.~\ref{fig:parspace}(d)) and hence an even 
larger neutralino relic density up to $\Omega_{\tz_1}h^2\sim 10000$. 
Points with $\Omega_{\tz_1}h^2\sim 0.1$ due to annihilation through Higgs can be found 
with both programs, but then the Yukawa unification is only $R\sim 1.07-1.09$\cite{bkss} 
while $f_H$ is of the order of 1\%. 
A plot of $\Omega_{\tz_1}h^2$ versus $R$ is shown in Fig.~\ref{fig:prelic}(b).

We conclude that 
both \Isajet\ and \Softsusy\ predict a large over-abundance of 
neutralino dark matter from Yukawa-unified models with $R\le 1.05$. 
The postulation of axion/axino dark matter allows one to reconcile the dark matter results
with Yukawa unification (as well as solving the strong $CP$ problem).

\FIGURE[t]{
\includegraphics[width=7cm]{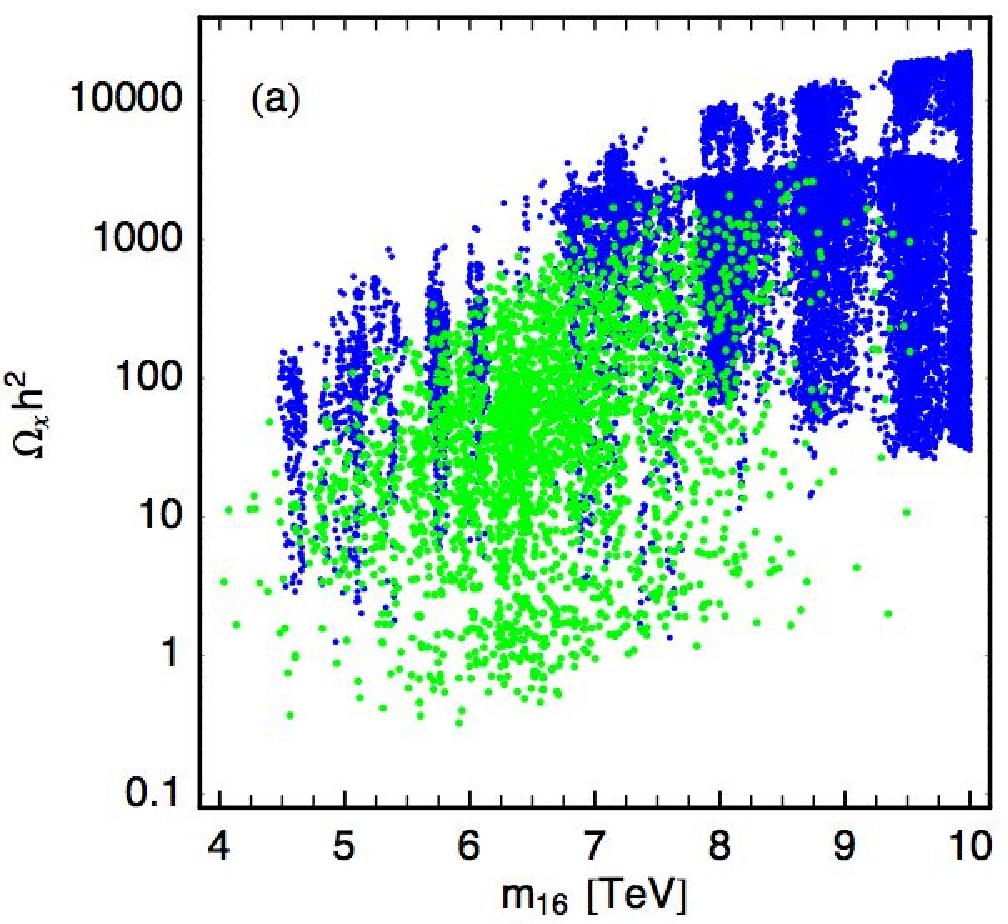}\quad
\includegraphics[width=7cm]{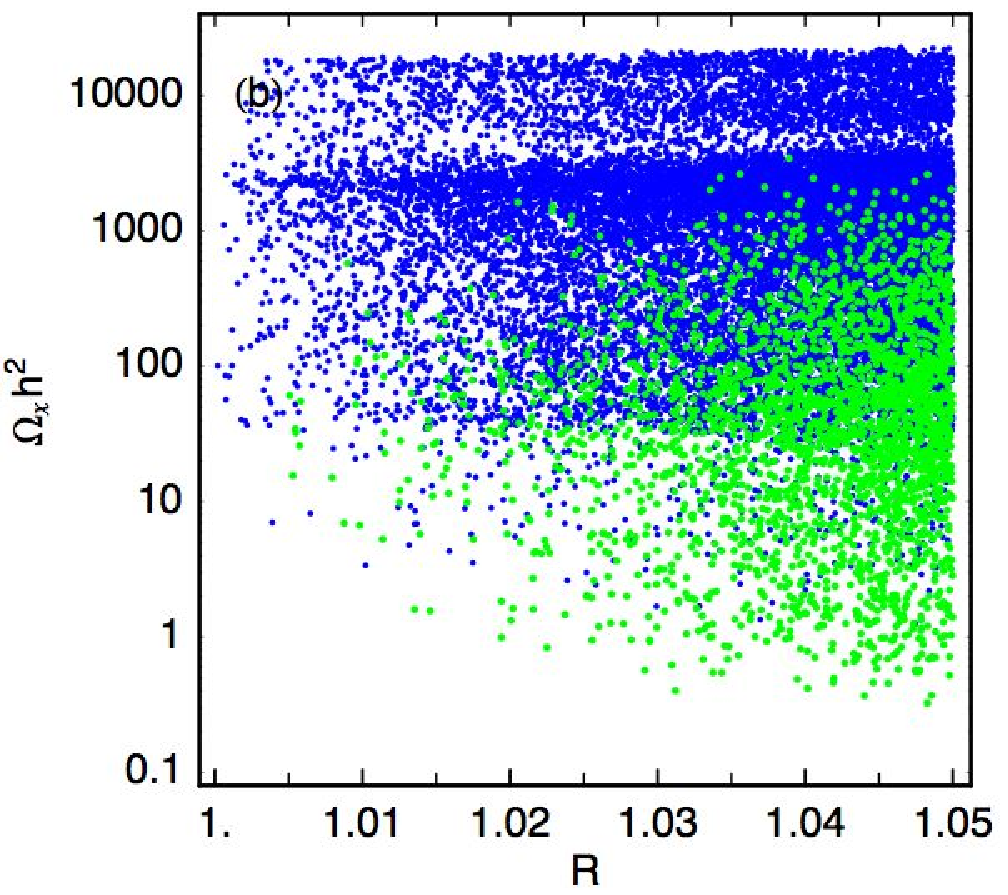}
\caption{Points with $R\le1.05$ from MCMC scans over the $SO(10)$ model parameter space,  
in (a)~$\Omega_{\tz_1}h^2$ vs.\ $m_{16}$ and
in (b)~$\Omega_{\tz_1}h^2$ vs.\ $R$.
\Isajet\,7.79 results are in blue, while \Softsusy\,2.0.18 results are in green.
}\label{fig:prelic}}

\subsection{Results for $b\to s\gamma$ and $B_s\to\mu^+\mu^-$ decay}
\label{ssec:bsg}

In this section, we present results for the branching fractions
for $b\to s\gamma $ and $B_s\to\mu^+\mu^-$ decays. 
We adopt the \Isajet\ Isatools\cite{isatools} program for these calculations.
The $BF(b\to s\gamma )$ decay rate is calculated in \Isajet\cite{bb} by evaluating the
Wilson co-efficients for the relevant operators mediating $b\to s\gamma$ 
decay at a scale $Q>M_{SUSY}$, and then running down to $M_Z$ scale
using a tower of effective theories approach\cite{anlauf}. At $M_Z$, the
Wilson co-efficients are matched to the SM ones, and run using 2-loop evolution
down to scale $m_b$, where the $b\to s\gamma$ decay rate is evaluated, including
complete NLO corrections\cite{greub}. Large $\tan\beta$ effects are accounted for
by adopting the \Isajet\ running Yukawa couplings, 
which include threshold effects as noted above\cite{bbct}.
The \Isajet\ SM result, $BF(b\to s\gamma )\simeq 3.1\times 10^{-4}$, agrees well
with a recent evaluation by Misiak\cite{misiak}, which finds 
$BF(b\to s\gamma )_{SM}=(3.15\pm 0.23)\times 10^{-4}$.

The branching fraction $BF(b\to s\gamma )$ has been measured by the
CLEO, Belle and BABAR collaborations; a combined analysis \cite{bsg_ex}
finds the branching fraction to be $BF(b\to s\gamma )=(3.55\pm
0.26)\times 10^{-4}$.
In Fig. \ref{fig:bsg}, we plot $BF(b\to s\gamma )$ for \Isajet\ Yukawa-unified solutions
with $R<1.05$ against $m_{16}$. The experimental central value along with $\pm 2\sigma$
error bands are also shown. The Wilson co-efficients grow with $\tan\beta$, and since
we are at $\tan\beta \sim 50$ for Yukawa-unified SUSY, we expect large MSSM contributions
to $BF(b\to s\gamma )$. In fact, we see from Fig. \ref{fig:bsg} that the 
$BF(b\to s\gamma )$ is typically {\it below} the measured error bands for $m_{16}\alt 10$ TeV,
and as $m_{16}$ gets very large, the branching fraction approaches the SM value.
This at least seems to favor $m_{16}$ values $\agt 8-10$ TeV in Yukawa-unified models.
This observation is also recorded by Altmannshofer et al.\cite{alt}.

\FIGURE[t]{
\epsfig{file=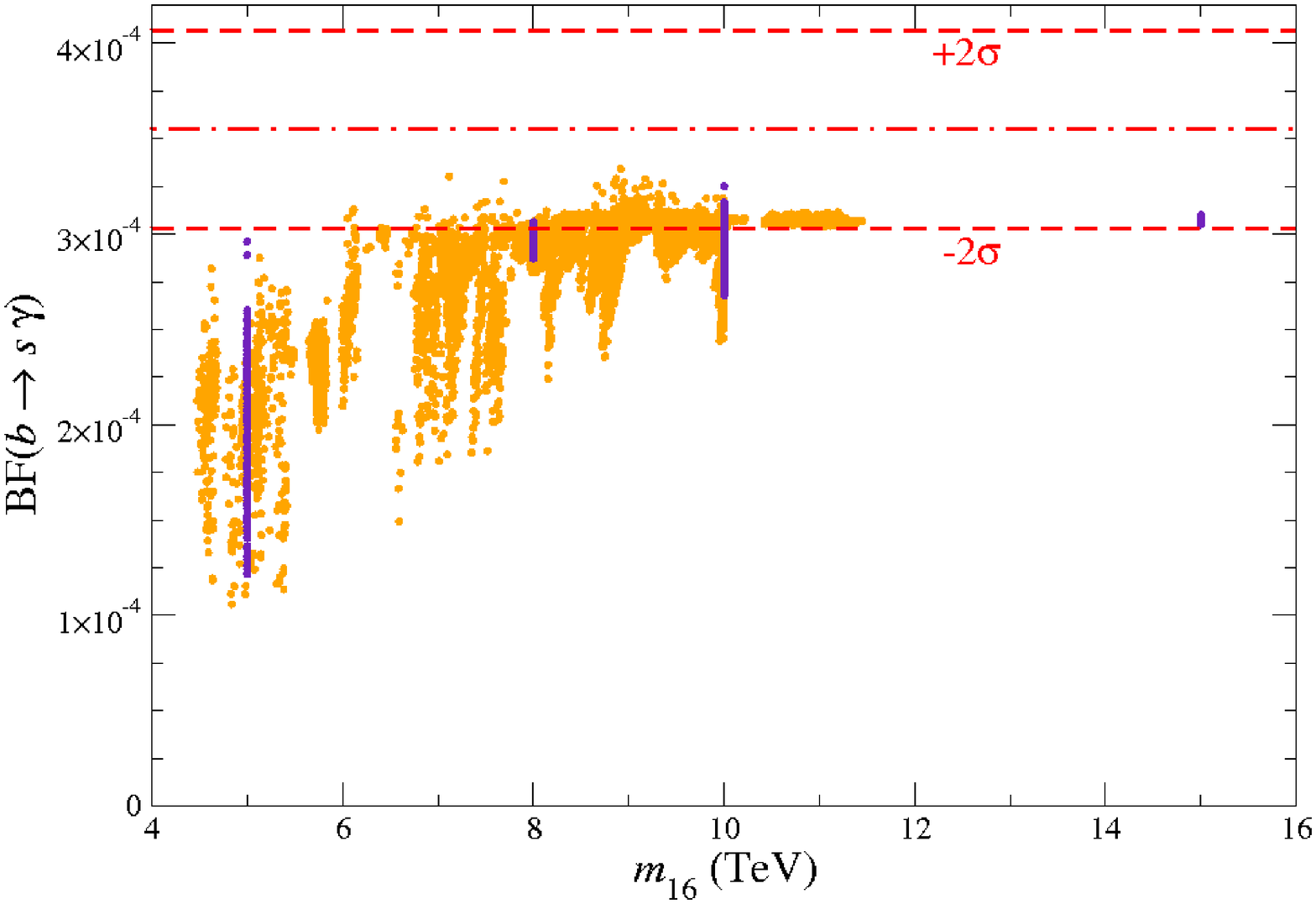,width=9cm,angle=0}
\caption{\label{fig:bsg} Predictions for $BF(b\to s\gamma )\ vs.\
m_{16}$ from Yukawa-unified models generated by \Isajet\,7.79 with
$R<1.05$ and $m_t=172.6$ GeV.
The blue dots show results from MCMC scans with fixed $m_{16}$.
We also show the measured central value of 
$BF(b\to s\gamma )$ along with $\pm 2\sigma$ error bands.}}  

\FIGURE[t]{
\epsfig{file=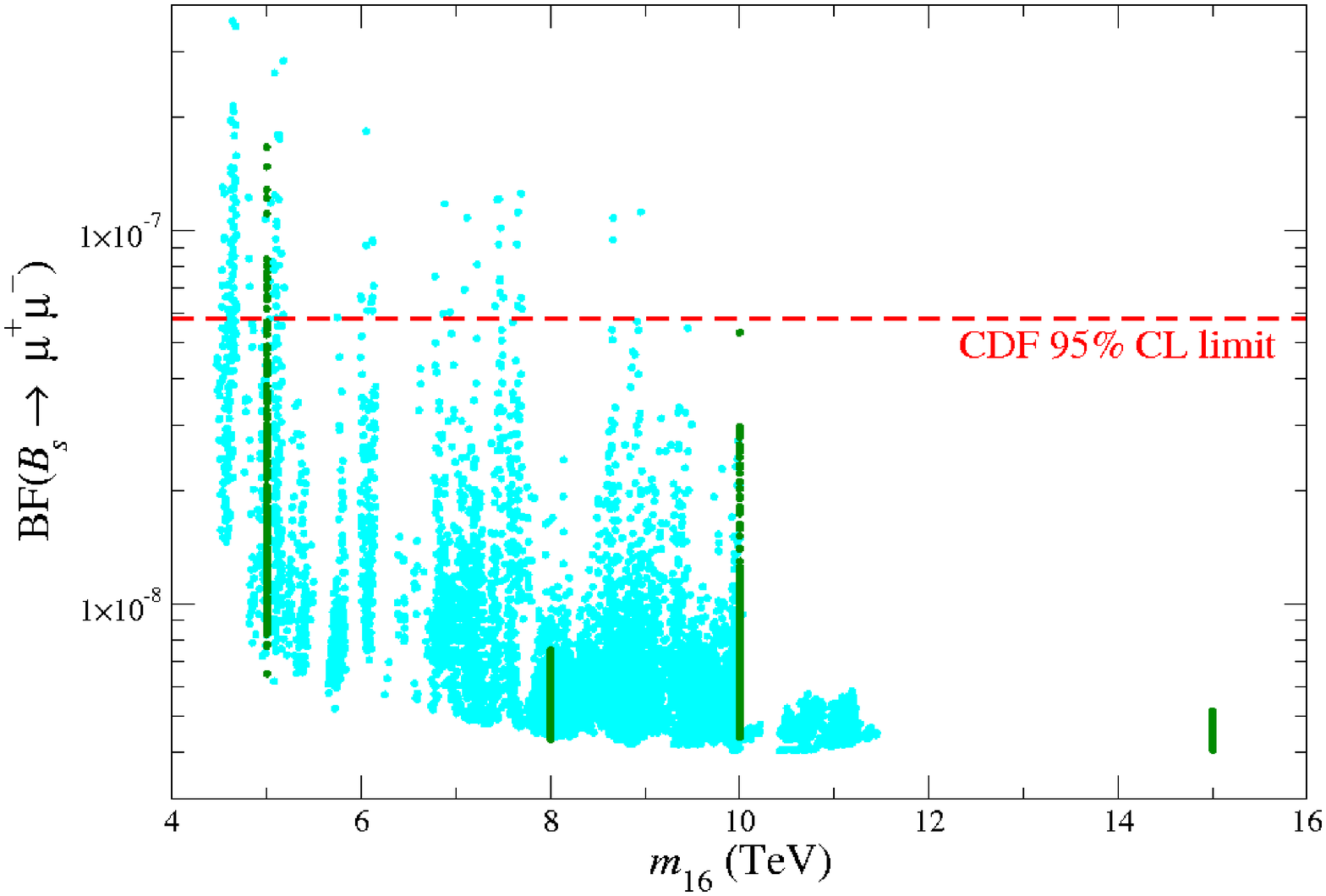,width=9cm,angle=0}
\caption{\label{fig:Bmm} Predictions for $BF(B_s\to\mu^+\mu^- )\ vs.\
m_{16}$ from Yukawa-unified models generated by \Isajet\,7.79 with
$R<1.05$ and $m_t=172.6$ GeV.
The green dots show results from MCMC scans with fixed $m_{16}$.
We also show the measured 95\% CL upper limit
on this branching fraction from CDF measurements.}}
 
In Fig. \ref{fig:Bmm}, we show the branching fraction $BF(B_S\to \mu^+\mu^- )$
evaluated\cite{tata} using Isatools. The experimental 95\% CL upper limit
from CDF collaboration is also shown\cite{cdf}. We see that for lower values of
$m_{16}$, in many cases the branching fraction can be near or even above 
the experimental limit. As $m_{16}$ increases, so does $m_A$, which mediates the decay.
The large value of $m_{16}$, and hence $m_A$, acts to suppress this branching fraction to
values about an order of magnitude below present experimental limits.

\section{Axion and axino dark matter}
\label{sec:dm}

\subsection{Axion cold dark matter}

The axion arises as a by-product of the Peccei-Quinn solution to the strong $CP$
problem\cite{pq,axreview}. The strong $CP$ problem has its origin in an allowed QCD Lagrangian term
\be
{\cal L}\ni \frac{\theta g^2}{32\pi^2}G_{\mu\nu}^a\tilde{G}^{a\mu\nu} 
\ee
($G_{\mu\nu}^a$is the gluon field strength tensor and $\tilde{G}^{a\mu\nu}$ its dual) which is $P$ and 
$T$-violating, but $C$ conserving, and hence $CP$ violating. 
When QCD is coupled to the electroweak theory, $\theta$ is replaced by 
$\bar{\theta}\equiv \theta +arg(det\ m_q )$, where $m_q$ is the quark mass matrix. 
The measured value of the neutron electric dipole moment (EDM) 
requires $\bar{\theta}\alt 10^{-10}$.
Explaining the tininess of this Lagrangian term is the strong $CP$ problem.

The Peccei-Quinn solution to the strong $CP$ problem invokes a theory with a global $U(1)$
(Peccei-Quinn or PQ) symmetry, which is classically valid, but broken spontaneously, and by
quantum anomalies. A consequence of the broken PQ symmetry is the existence of a 
pseudo-Goldstone boson field: the axion $a(x)$\cite{ww}. 
In this case, the Lagrangian also contains the terms
\be
{\cal L}\ni {1\over 2}\partial_\mu a\partial^\mu a +\frac{g^2}{32\pi^2}\frac{a(x)}{f_a/N}
G_{\mu\nu}^a \tilde{G}^{a\mu\nu} , \label{L_ax}
\ee
where we have introduced the PQ breaking scale $f_a$ and $N$ is the model-dependent color anomaly
of order 1.
The effective potential for the axion field $V(a(x))$ has its minimum at 
$\langle a(x)\rangle = -\bar{\theta} f_a/N$, and so the offending $G\tilde{G}$ term
essentially vanishes, which solves the strong $CP$ problem. 
A consequence of this very elegant mechanism is that
a physical axion field should exist, with concommitant particle excitations.

The axion mass can be computed using current algebra techniques, and is given by
\be
m_a\simeq 6\ {\rm eV}\frac{10^6\ {\rm GeV}}{f_a/N} .
\label{eq:axmass}
\ee
The axion field couples to gluon-gluon (obvious from Eq. (\ref{L_ax})) and also to photon-photon 
and fermion-fermion. All the couplings are suppressed by the PQ scale $f_a$. 

Astrophysical limits
from cooling of red giant stars and supernova 1987a require $f_a/N \agt 10^9$ GeV, or
$m_a\alt 3\times 10^{-3}$ eV. In addition, axions can be produced via various mechanisms
in the early universe. Since their lifetime (they decay via $a\to\gamma\gamma$) turns out to
be longer than the age of the universe, they can be a good candidate for dark matter in the universe. 
Since we will be concerned here with re-heat temperatures of the universe $T_R\alt 10^9\ {\rm GeV}<f_a$
(to avoid overproducing gravitinos in the early universe), the axion production mechanism relevant
for us here is just one: production via vacuum mis-alignment\cite{absik}. In this mechanism, the axion field
$a(x)$ can have any value $\sim f_a$ at temperatures $T\gg \Lambda_{QCD}$. As the temperature 
of the universe drops,
the potential turns on, and the axion field oscillates and settles to its minimum at $-\bar{\theta} f_a/N$.
The difference in axion field before and after potential turn-on corresponds to 
the vacuum mis-alignment: it produces an axion number density
\be
n_a(t)\sim {1\over 2}m_a(t)\langle a^2(t)\rangle ,
\ee
where $t$ is  the time near the QCD phase transition.
Relating the number density to the entropy density allows one to determine the
axion relic density today:
\be
\Omega_a h^2\simeq {1\over 4}\left(\frac{6\times 10^{-6}\ {\rm eV}}{m_a}\right)^{7/6} .
\label{eq:axrelic}
\ee
An error estimate of the axion relic density from vacuum mis-alignment is 
plus-or-minus a factor of three. 
Axions produced via vacuum mis-alignment would constititute {\it cold} dark matter.

The axion relic density from vacuum mis-alignment, along with error bands, is shown 
in Fig. \ref{fig:axion}. However, in the event that $\langle a^2(t)\rangle$ is 
inadvertently small, then much lower values of relic density could be allowed (or much higher if
$\langle a^2(t)\rangle$ is inadvertently large).
Additional entropy production at $t>t_{QCD}$ can also lower the axion relic abundance.
Taking the value of Eq.~(\ref{eq:axrelic}) literally, 
and comparing to the WMAP5 measured abundance of CDM in the universe, 
one gets an upper bound $f_a/N\alt 5\times 10^{11}$ GeV, or a lower bound
$m_a\agt 10^{-5}$ eV. If we take the axion
relic density a factor of three lower, then the bounds change
to $f_a/N \alt 1.2\times 10^{12}$ GeV, and $m_a\agt 4\times 10^{-6}$ eV. 

\FIGURE[t]{
\epsfig{file=vacalign.eps,width=10cm} 
\caption{Axion relic density due to vacuum mis-alignment
versus $m_a$ (lower scale) and $f_a/N$ (upper scale). The plot
includes a factor three error estimate along with the WMAP5 CDM measured abundance.
}\label{fig:axion}}
%

\subsection{Warm and cold axino dark matter}

\subsubsection{Non-thermally produced axino dark matter}

Since we are working in a supersymmetric model, the axion field will be only one element
of an axion left chiral scalar superfield 
\be
\hat{\phi}_a = \frac{s(\hat{x})+ia(\hat{x})}{\sqrt{2}}+i\sqrt{2}\bar{\theta}\psi_{a_L}(\hat{x})+i\bar{\theta}
\theta_L{\cal F}_a(\hat{x} ) ,
\ee
where $\theta$ here are the anti-commuting Grassman superspace dimensions arranged in a Majorana
spinor, and $\hat{x}_\mu =x_\mu+{i\over 2}\bar{\theta}\gamma_5\gamma_\mu\theta$\cite{wss}.
The superfield $\hat{\phi}_a$ contains the $R$-even spin-0 saxion field $s$, which gets a mass of order the SUSY
breaking scale, and the $R$-odd spin-$1\over 2$ axino field $\psi_a\equiv\ta$, whose mass is model-dependent, and
can range over the keV-GeV scale\cite{axmass,ckkr}. 

Here, we assume that the $\ta$ is the LSP, so that 
the neutralino is in fact unstable, and decays dominantly into $\tz_1\to \ta\gamma$. 
The width $\Gamma (\tz_1\to\ta\gamma )$ has been calculated in Ref. \cite{ckkr}, 
and is given by
\be
\Gamma (\tz_1\to\ta\gamma )=\frac{\alpha_{em}^2 C_{aYY}v_4^{(1)2}}{128\pi^3\cos^2\theta_W}
\frac{m_{\tz_1}^3}{(f_a/N)^2}\left( 1-\frac{m_{\ta}^2}{m_{\tz_1}^2}\right)^3 ,
\ee
\label{eq:Gaxino}
where $v_4^{(1)}$ denotes the bino fraction of neutralino $\tz_1$, $N$ is the 
model-dependent anomaly factor ({\it e.g.} $N=1\ (6)$ for KSVZ\cite{ksvz} (DFSZ\cite{dfsz}) axions),
and $C_{aYY}$ is a model-dependent coupling factor ({\it e.g.} $C_{aYY}=8/3$ in
the DFSZ model).

In Fig. \ref{fig:tau_axino}, we plot the $\tz_1$ lifetime in seconds versus $m_{\tz_1}$
for four choices of $f_a/N$, and taking $C_{aYY}=8/3$. 
The lifetime ranges from $\sim 10^{-5}$s for $f_a/N =10^9$ GeV, up to $\sim 40$ s
for $f_a/N=10^{12}$ GeV. In the latter case, the $\tz_1$ will decay while BBN is ongoing.
The dominant decay into a high energy photon should thermalize with the electron-nucleon plasma.
It is of note that $\tz_1\to q\bar{q}\ta$ three body hadronic decays via intermediate
$\gamma$ and $Z$ can also occur at a small branching fraction. These hadronic decays
would be more likely to be a threat to disrupt Big Bang Nucleosynthesis. 
Also, $\tz_1\to Z\ta$ can occur, but only for $m_{\tz_1}>M_Z$, which rarely occurs
in our Yukawa-unified scenario.

\FIGURE[t]{
\epsfig{file=life.eps,width=10cm} 
\caption{Lifetime in seconds of the lightest neutralino versus its mass 
for $f_a/N=10^9$, $10^{10}$, $10^{11}$ and $10^{12}$ GeV, respectively.
}\label{fig:tau_axino}}

The axino dark matter produced from neutralino decay would compose 
non-thermally produced (NTP) dark matter. Jedamzik {\it et al.}\cite{jlm} have calculated
the rms velocity profile of axino dark matter coming from neutralino decay.
A comparison against data from Lyman alpha forest leads them to conclude that
non-thermally produced axinos will consitute warm dark matter for $m_{\ta}\alt 1$ GeV.

The relic abundance of non-thermally produced axinos can be simply obtained from the
neutralino abundance. The neutralino thermal abundance calculation proceeds by solving the
Boltzmann equation for neutralinos from freeze-out to the present day, after inputting the
usual neutralino annihilation and co-annihilation cross sections. Since each neutralino
decays to one axino, the axinos inherit the neutralino number density, and the
non-thermally produced axino abundance is simply
\be
\Omega_{\ta}^{\rm NTP}h^2 =\frac{m_{\ta}}{m_{\tz_1}}\Omega_{\tz_1}h^2 .
\label{eq:axino_ntp}
\ee
In this regard, if the ratio $m_{\ta}/m_{\tz_1}$ is small, then large factors of
neutralino dark matter density can be shed by undergoing $\tz_1\to\ta\gamma$ decay. It is this 
mechanism that allows one to reconcile the huge neutralino relic abundance from
Yukawa-unified models with the WMAP measured abundance.

\subsubsection{Thermally produced axino dark matter}

In our scenario, where we only consider $T_R\ll f_a$, the axinos in the
early universe are too weakly interacting to be in thermal equilibrium. Nevertheless,
they can be produced by radiation off other particles which are in the thermal bath, much
the same as gravitinos can be produced in the early universe. 
Initial calculations of the thermally produced (TP) axino abundance were performed in 
Ref. \cite{ckkr}, wherein a variety of QCD axino production processes (such as
$gg\to\ta\tg$, $g\tq\to\ta q$, $\cdots$) were considered. Divergent diagrams
involving $t$-channel exchange of massless gluons were regulated by introducing a 
``plasmon'' mass, representing the effective gluon mass in the plasma of the early universe.
A later evaluation of thermally produced axino matter in Ref. \cite{steffen} used the hard thermal loop 
resummation technique of Braaten-Pisarski\cite{braaten} and obtained a reduced axino dark matter yield 
by a factor of $\sim 3$. The results of Ref. \cite{steffen} are summarized in the expression
\be
\Omega_{\ta}^{\rm TP}h^2\simeq 5.5 g_s^6\ln\left(\frac{1.108}{g_s}\right)
\left(\frac{10^{11}\ {\rm GeV}}{f_a/N}\right)^2 
\left(\frac{m_{\ta}}{0.1\ {\rm GeV}}\right) 
\left(\frac{T_R}{10^4\ {\rm GeV}}\right)
\label{eq:axino_tp}
\ee
where $g_s$ is the strong coupling evaluated at $Q=T_R$ 
({\it e.g.} $g_s =.915$ at $Q=10^6$ GeV from our \Isajet\ RGE calculations). 
The thermally produced axinos qualify as {\it cold} dark matter as long as 
$m_{\ta}\agt 100$ keV\cite{ckkr,steffen}.

In Fig. \ref{fig:axino_rd1}, we show bands of the $m_{\ta}\ vs.\ T_R$ plane
which give $\Omega_{\ta}^{\rm TP}h^2$ within the WMAP-measured dark matter abundance,
for $f_a/N=10^{10}$, $10^{11}$ and $10^{12}$ GeV. We see that for the lower
range of $f_a/N\sim 10^{10}$ GeV, very low values of $m_{\ta}$ and $T_R$ are
required. In this case, with $m_{\ta}\alt 100$ keV, the thermally produced
axino DM would likely constitute warm DM, and furthermore, the low value of 
$T_R$ excludes some of the possible mechanisms for baryogenesis. In this case, if
we want $T_R\agt 10^6$ GeV and $m_{\ta}\agt 100$ keV with dominant thermal
production of cold axino DM, then we will need higher $f_a/N\agt 10^{11}$ GeV.

\FIGURE[t]{
\epsfig{file=mavtr.eps,width=10cm} 
\caption{Thermally produced axino relic density 
within the WMAP5 measured limits in the $m_{\ta}$ vs. $T_R$
plane for $f_a/N=10^{10}$, $10^{11}$ and $10^{12}$ GeV, 
respectively.
}\label{fig:axino_rd1}}

In Fig. \ref{fig:axino_rd2}, we show bands of 
$\Omega_{\ta}^{\rm TP}h^2=0.11,\ 0.03,\ 0.01$ and $0.001$ in the 
$m_{\ta}\ vs.\ T_R$ plane for $f_a/N= 10^{12}$ GeV. In this case, if
the thermally produced axino dark matter only constitutes a small fraction
of the total dark matter, and most of the remainder is composed of cold axions, 
then much smaller values of $m_{\ta}$ are allowed, and the thermally produced
axinos can be either warm or even hot dark matter.

\FIGURE[t]{
\epsfig{file=omghmavtr.eps,width=10cm} 
\caption{Thermally produced axino relic density 
in the $m_{\ta}$ vs. $T_R$ plane for $f_a/N=10^{12}$ GeV.
}\label{fig:axino_rd2}}
%

\section{The gravitino problem, non-thermal leptogenesis and mixed axion/axino dark matter
\label{sec:axdm}}

\subsection{The gravitino problem}
\label{ssec:gravitino}

A problem common to all 
SUSY models including supergravity (SUGRA) is known as the gravitino problem.
In realistic SUGRA models (those that include the SM as their sub-weak-scale effective
theory), SUGRA is broken in a hidden sector by the superHiggs mechanism.
A mass for the gravitino $\tG$ is induced by SUGRA breaking, 
which is commonly taken to be of order the weak scale. 
The gravitino mass $m_{3/2}$ sets the mass scale
for all the soft breaking terms, so that all SSB terms end up also being of order
the weak scale\cite{cremmer,nilles}.

The coupling of the gravitino to matter is strongly suppressed by the Planck 
mass, so the $\tG$ in the mass range considered here ($m_{3/2}\sim m_{16}\sim 5-20$ TeV) 
is never in thermal equilibrium with the thermal bath in 
the early universe. Nonetheless, it does get produced by scatterings of
particles that do partake of thermal equilibrium.
Thermal production of gravitinos in the early universe has been calculated 
in Refs. \cite{relic_G}, where the abundance is found to depend naturally on 
$m_{3/2}$ and on the re-heat temperature $T_R$ at the end of inflation.
Once produced, the $\tG$s decay into all varieties of particle-sparticle
pairs, but with a lifetime that can exceed $\sim 1$ sec, the time scale
where Big Bang nucleosynthesis (BBN) begins. 
The energy injection from $\tG$ decays is a threat to dis-associate the
light element nuclei which are created in BBN.
Thus, the long-lived $\tG$s can destroy the successful predictions of the 
light element abundances as calculated by nuclear thermodynamics.

The BBN constraints on gravitino production in the early universe have been
calculated by several groups\cite{bbn_gino}. The recent 
results from Ref. \cite{kohri} give an upper limit on the re-heat temperature
as a function of $m_{3/2}$. The results depend on how long-lived the $\tG$
is (at what stage of BBN the energy is injected), and what its dominant decay
modes are. Qualitatively, for $m_{3/2}\alt 5$ TeV, $T_R\alt 10^6$ GeV
is required; if this is violated, then too many $\tG$ are produced in the early
universe, which detroy the $^3He$, $^6Li$ and $D$ abundance calculations.
For $m_{3/2}\sim 5-50$ TeV, the re-heat upper bound is much less:
$T_R\alt 10^9$ GeV (depending on the $^4He$ abundance) 
due to overproduction of $^4He$ arising from
$n\leftrightarrow p$ conversions. For $m_{3/2}\agt 50$ TeV, there is an upper bound 
of $T_R\alt 10^{10}$ GeV due to overproduction of $\tz_1$ LSPs due to 
$\tG$ decays. 

Solutions to the gravitino BBN problem then include: 1. having $m_{3/2}\agt 50$ TeV
but with an unstable $\tz_1$ ($R$-parity violation and no $T_R$ bound),
2. having a gravitino LSP so that $\tG$ is stable or 3. keep the re-heat
temperature below the BBN bounds. We will here adopt solution number 3. 
In the case of $SO(10)$ SUSY GUT models, we expect $m_{3/2}\sim m_{16}$ and since $m_{16}\sim 5-20$ TeV, 
this means we need a re-heat temperature $T_R\alt 10^9$ GeV. 

\subsection{Non-thermal leptogenesis}

The data gleaned on neutrino masses during the past decade has led credence to 
a particular mechanism of generating the baryon asymmetry of the universe 
known as leptogenesis\cite{leptog}. Leptogenesis requires the presence of heavy 
right-handed 
gauge-singlet Majorana neutrino states $\psi_{N^c_i}(\equiv N_i)$ with mass $M_{N_i}$
(where $i=1-3$ is a generation index). 
The $N_i$ states may be produced thermally in the early universe, or perhaps 
non-thermally, as suggested in Ref. \cite{NTlepto} via inflaton $\phi \to N_iN_i$
decay. The $N_i$ may then decay asymmetrically to elements of the doublets-- 
for instance $\Gamma (N_1\to h_u^+ e^-)\ne \Gamma (N_1\to h_u^- e^+)$-- owing
to the contribution of $CP$ violating phases in the tree/loop decay
interference terms. Focusing on just one species of heavy neutrino $N_1$, 
the asymmetry is calculated to be\cite{epsilon}
\be
\epsilon\equiv \frac{\Gamma (N_1\to \ell^+)-\Gamma (N_1\to \ell^-)}{\Gamma_{N_1}}
\simeq -\frac{3}{8\pi}\frac{M_{N_1}}{v_u^2}m_{\nu_3}\delta_{\rm eff} ,
\ee 
where $m_{\nu_3}$ is the heaviest active neutrino, $v_u$ is the up-Higgs vev and
$\delta_{\rm eff}$ is an effective $CP$-violating phase factor which may be of order 1.
The ultimate baryon asymmetry of the universe is proportional to $\epsilon$, 
so larger values of $M_{N_1}$ lead to a higher baryon asymmetry. 

To find the baryon asymmetry, one may first assume that the $N_1$ is thermally 
produced in the early universe, and then solve the Boltzmann equations for the $B-L$
asymmetry. The ultimate baryon asymmetry of the universe arises from the lepton asymmetry
via sphaleron effects. The final answer\cite{bp}, compared against 
the WMAP-measured result $\frac{n_B}{s}\simeq 0.9\times 10^{-10}$ for the baryon-to-entropy
ratio, requires $M_{N_1}\agt 10^{10}$ GeV, and thus a re-heat temperature $T_R\agt 10^{10}$ GeV.
This high a value of reheat temperature is in conflict with the upper bound on $T_R$
discussed in Sec.~\ref{ssec:gravitino}. 
In this way, it is found that generic SUGRA models with $R$-parity conservation are apparently 
in conflict with thermal leptogenesis as a means to generate the baryon asymmetry of the universe.

If one instead looks to non-thermal (NT) leptogenesis, then it is possible to have lower
reheat temperatures, since the $N_1$ may be generated via inflaton decay. 
The Boltzmann equations for the $B-L$ asymmetry have been solved numerically in Ref. \cite{imy}.
The $B-L$ asymmetry is then converted to a baryon asymmetry via sphaleron effects
as usual. 
The baryon-to-entropy ratio is found to be\cite{imy}
\be
\frac{n_B}{s}\simeq 8.2\times 10^{-11}\times \left(\frac{T_R}{10^6\ {\rm GeV}}\right) 
\left(\frac{2M_{N_1}}{m_\phi}\right) \left(\frac{m_{\nu_3}}{0.05\ {\rm eV}}\right) \delta_{eff} ,
\ee
where $m_\phi$ is the inflaton mass.
Comparing calculation with data, a lower bound $T_R\agt 10^6$ GeV may be inferred
for viable non-thermal leptogenesis via inflaton decay.

\subsection{A consistent cosmology for Yukawa-unified models with mixed axion/axino dark matter}
\label{ssec:results}

Next, we investigate various scenarios with mixed axion/axino cold and warm dark matter,
first to see if they can yield a consistent cosmology, and second, to see if cosmology
provides any insight into allowed model parameters.
Here, we will investigate four cases.
\begin{enumerate}
\item We take $f_a/N = 10^{11}$ GeV. Using the central value from Eq.~\ref{eq:axrelic}, we obtain 
a small fraction of axion CDM: $\Omega_a h^2\simeq 0.017$. 
The bulk of CDM must then be composed of something else: in our case,
thermally produced axinos (so $m_{\ta}\agt 100$ keV). We take $\Omega_{\ta}^{\rm TP}h^2=0.083$.
Then to obtain the WMAP5 measured DM abundance, we get $\Omega_{\ta}^{\rm NTP}h^2\sim 0.01$.
\item We take $f_a/N =4\times 10^{11}$ GeV so that the central value of axion relic abundance 
Eq. \ref{eq:axrelic} yields $\Omega_a h^2\simeq 0.084$. This gives dominant axion CDM, so that
thermally and non-thermally produced axino abundance should be small. Here we assume an equal
mix of thermal and non-thermal axinos, so 
$\Omega_{\ta}^{\rm TP}h^2=\Omega_{\ta}^{\rm NTP}h^2=0.013$.
\item We take $f_a/N = 10^{12}$ GeV, and a factor of $1/3$ error on $\Omega_a h^2$
as in the lower dashed curve of Fig. \ref{fig:axion}. Thus, $\Omega_a h^2=0.084$ so we
have dominant axion CDM. As in the previous case we take an equal mix of thermally and 
non-thermally produced axinos: $\Omega_{\ta}^{TP}h^2=\Omega_{\ta}^{NTP}h^2=0.013$.
\item Here, we again take $f_a/N = 10^{12}$ GeV, but assume the axion vev is 
accidentally close to zero so that it is nearly aligned with the 
potential minimum, instead of mis-aligned. Even though $f_a/N$ is large, the resulting axion 
abundance is small: $\Omega_a h^2\sim 0$. In this case, thermally produced axinos should 
make up the dominant CDM component. 
We take $\Omega_{\ta}^{\rm TP}h^2=0.1$ and $\Omega_{\ta}^{\rm NTP}h^2\sim 0.01$.
This case was shown previously as one adopted in Ref. \cite{bs}.
\end{enumerate}

Once the value of $\Omega_{\ta}^{\rm NTP}h^2$ is known, we may calculate $\Omega_{\tz_1}h^2$ and
$m_{\tz_1}$ in any supersymmetric model (with a $\tz_1$ NLSP) and use Eq.~(\ref{eq:axino_ntp}) 
to calculate the value of $m_{\ta}$ that is needed. Then, if a value of $f_a/N$ has been selected, 
and we know $\Omega_{\ta}^{\rm TP}h^2$, we can use Eq.~(\ref{eq:axino_tp}) to determine the 
required re-heat temperature $T_R$. We plot our final results in the $m_{\ta}\ vs.\ T_R$ plane, 
so that we may see whether a consistent cosmological scenario may be found for
any of our Yukawa-unified solutions, and also whether a consistent cosmology helps to select out
preferred values of the soft SUSY breaking parameters.

Our main results are shown in Fig. \ref{fig:maxvsTR}. Here, we generate Yukawa-unified solutions
with $R\le1.05$ first with \Isajet\,7.79 through 
MCMC scans for $m_{16}=5$ TeV (small-red points), 8 TeV (dark-blue larger points), 
10 TeV (medium-blue larger points) and 15 TeV (very large light-blue points).
The cases 1--4 are labelled
as C1--C4. For the region below $T_R\sim 10^4$ GeV, the calculation of thermally produced
axinos breaks down. Moreover, the value of $T_R$ is becoming comparable to the sparticle mass scale
({\it i.e.} the heavier sparticles will not be produced in thermal equilibrium in the early universe) so the
calculation of $\Omega_{\tz_1}h^2$ would also break down. Values of $T_R\agt 10^6$ are
compatible with non-thermal leptogenesis via inflaton decay.

\FIGURE[t]{
\epsfig{file=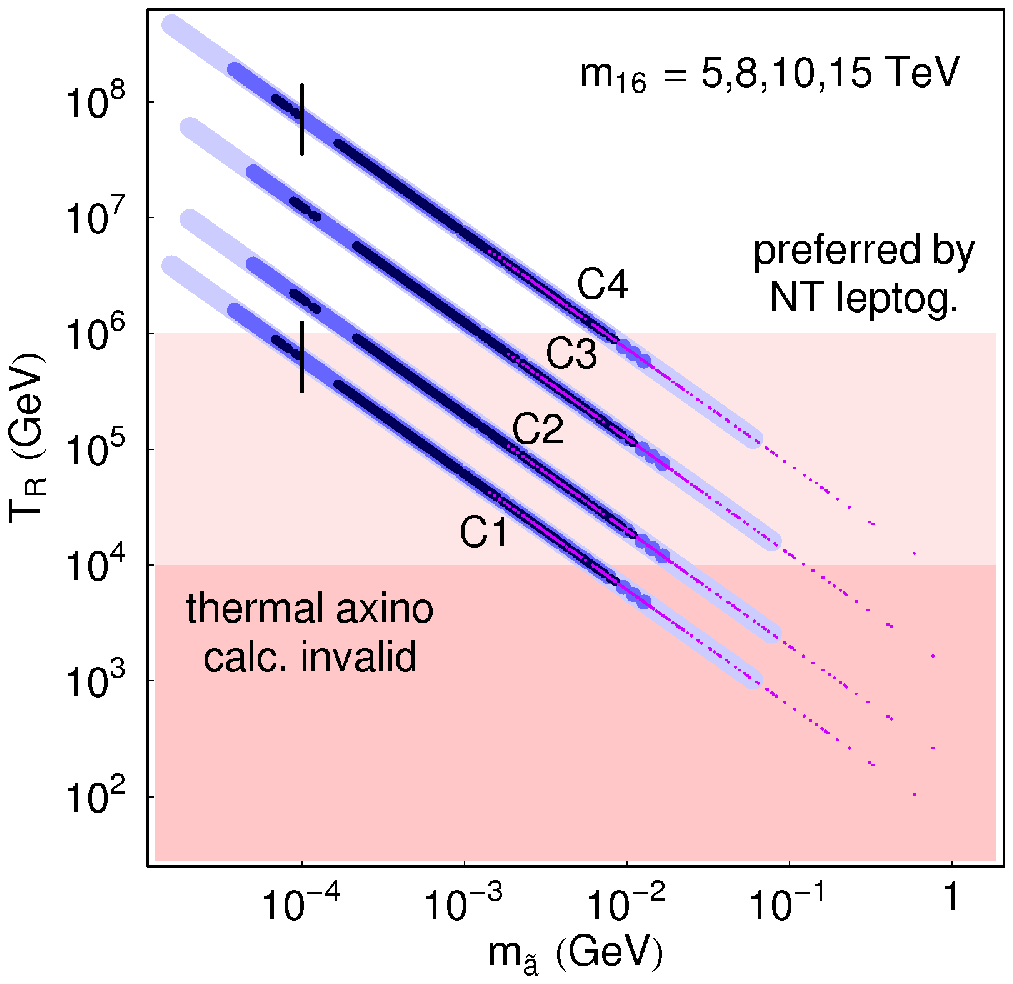,width=10cm} 
\caption{Plot of locus in the $m_{\ta}\ vs.\ T_R$ plane of
four Yukawa-unified cases of mixed axion/axino dark matter, along with
four different $m_{16}$ values. The thermally produced axino relic density 
calculation is only valid for $T_R\agt 10^4$ GeV. Non-thermal leptogenesis
prefers the region with $T_R\agt 10^6$ GeV.
}\label{fig:maxvsTR}}

\begin{itemize}
\item We see from Fig. \ref{fig:maxvsTR} that case C1 with $f_a/N=10^{11}$ GeV 
and dominantly TP axino dark matter gives almost all
solutions in the region with $T_R<10^6$ GeV. The solutions with $m_{16}\sim 5-8$ TeV
especially have low values of $T_R$. The solutions with $m_{16}\sim 15$ TeV do have 
$T_R\agt 10^6$ GeV, but these solutions also have $m_{\ta}<10^{-4}$ GeV, and so the dominant
DM component from thermally produced axinos is likely warm DM. This scenario would thus be difficult
to accept cosmologically, for any value of $m_{16}$.
\item For case C2, we have $f_a/N=4\times 10^{11}$ GeV and dominant axion CDM. Here, the larger
value of $f_a/N$ allows solutions with a similar value of $m_{\ta}$ as case C1, but with a 
higher value of $T_R$. The solutions with $m_{16}=10-15$ TeV do emerge with $T_R\agt 10^6$ GeV--
in the range for NT leptogenesis. Many of these solutions still have $m_{\ta}<10^{-4}$ GeV,
so that the thermally produced axinos are warm. In this case, $m_{\ta}<10^{-4}$ GeV is allowed, 
since instead the axions actually make up the CDM.
\item In case C3, $f_a/N= 10^{12}$ GeV with dominant axion CDM. The larger $f_a/N$ gets, the larger are the 
calculated values of $T_R$. 
The thermal and non-thermal axinos both have small contributions to the relic density,
so the entire band of solutions with $T_R>10^6$ GeV yields a consistent cosmology. This case requires
an axino with $m_{\ta}\alt 10^{-3}$ GeV for $T_R>10^6$ GeV.
\item Finally, case C4 is constructed to have a large value of $f_a/N=10^{12}$ GeV, but with a tiny
axion relic abundance due to accidental vacuum alignment. In this case, the thermally produced
axinos comprise the CDM. Solutions are found with $T_R>10^6$ GeV for $m_{\ta}\alt 6\times 10^{-3}$ GeV. 
However, in this case,
the solutions with $m_{\ta}\alt 10^{-4}$ GeV would not be allowed, since they likely yield
a dominant warm DM scenario, in contrast to requirements from large scale structure formation that the bulk
of DM be cold.
\end{itemize}

\FIGURE[t]{
\epsfig{file=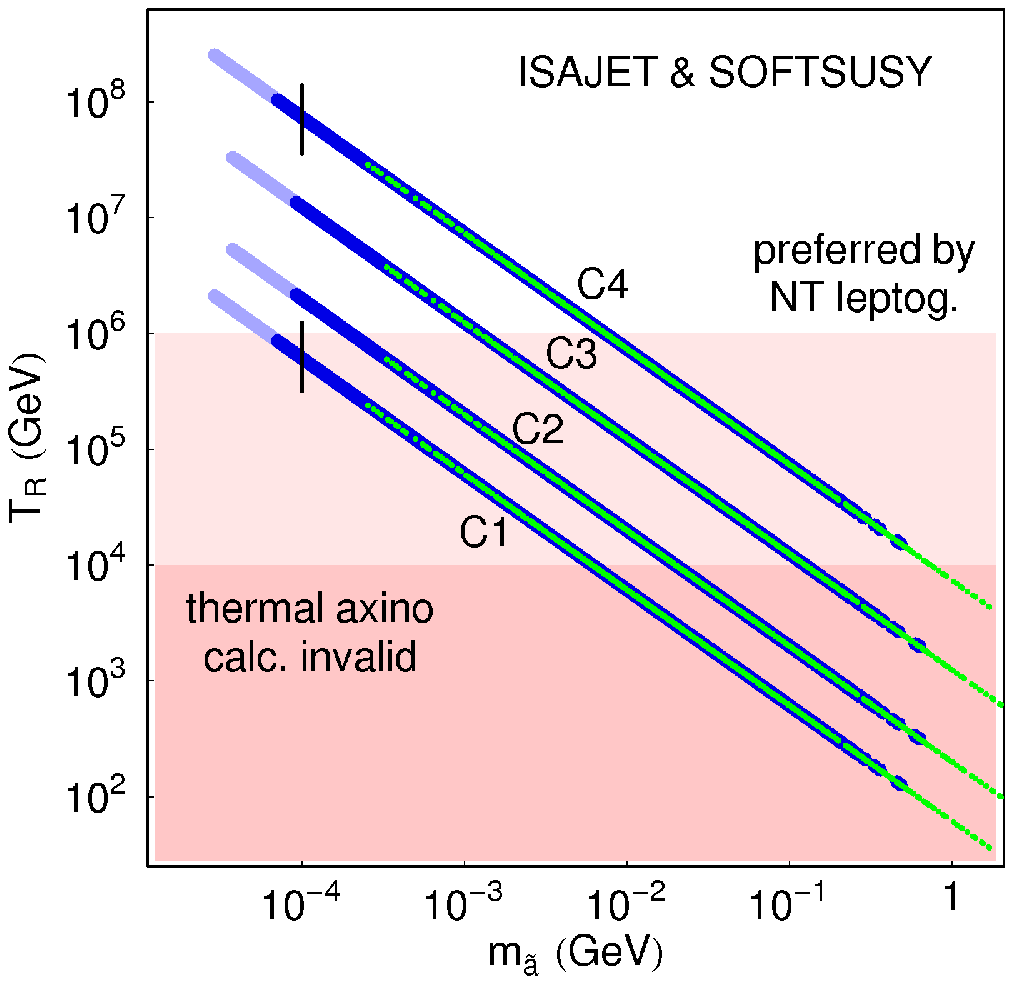,width=10cm} 
\caption{ Plot of locus in the $m_{\ta}\ vs.\ T_R$ plane of
four Yukawa-unified cases of mixed axion/axino dark matter for all 
$m_{16}$ values. We show results from \Isajet\ (blue) and \Softsusy\ (green).
}\label{fig:maxvsTR_softs}}

In Fig. \ref{fig:maxvsTR_softs}, we show Yukawa-unified solutions 
with $R<1.05$ for cases C1--C4 in the $m_{\ta}\ vs.\ T_R$ plane from 
both \Isajet\,7.79 and \Softsusy\,2.0.18  MCMC scans.\footnote{To be precise, 
the results for $m_{16}=5$, 8, and 10~TeV come from both \Isajet\,7.79 and \Softsusy\,2.0.18.
Points with $m_{16}=15$~TeV come only from  \Isajet\,7.79, because \Softsusy\,2.0.18 does not 
give consistent EWSB for such high $m_{16}$.} \Isajet\ results are in blue, while \Softsusy results are 
in green. Both sets of results line on the same line for a given case. The \Isajet\ results
reach to somewhat higher $T_R$ values than \Softsusy. This is due in part because
\Isajet\ can more easily generate Yukawa-unified models for very high $m_{16}\agt 10$ TeV,
and these models typically have larger $\Omega_{\tz_1}h^2$ values, and hence 
smaller values of $m_{\ta}$. Also, the \Softsusy\ results tend to have lower $\mu$ values 
than \Isajet\ results, which also tends to lower the value of the 
neutralino relic density. Thus, case C1 does not lead to a consistent cosmology for either 
\Isajet\ or \Softsusy. \Isajet\ can obtain cosmologies with $T_R>10^6$ GeV for cases C2--C4, but \Softsusy\
can generate cosmologies with $T_R>10^6$ GeV only for case C3 (just barely) and case C4.

\section{Summary and conclusions}
\label{sec:conclude}

One vestige of supersymmetric $SO(10)$ grand unified theories may be
that the third generation $t-b-\tau$ Yukawa couplings unify, in
addition to gauge couplings and matter multiplets.
Assuming the MSSM is the low energy effective theory at energy
scales $Q<M_{GUT}$, we are able to find parameter space solutions that
yield a superparticle mass spectrum with Yukawa coupling unification
good to 5\% or better, using the \Isajet\,7.79 and \Softsusy\,2.0.18 programs.
The sparticle mass spectrum that qualitatively emerges is that
first and second generation scalars lie in the multi-TeV regime,
third generation scalars, $\mu$ and $m_A$ lie in the few TeV range, 
and gauginos lie in the sub-TeV range. 
The neutralino relic density turns out to be $10-10^4$ times the measured
dark matter density, prompting the suggestion that in this case,
the axino is a better LSP candidate, so that the dark matter of the universe
would be composed of an axion/axino mix. 

Yukawa-unified SUSY models should thus give rise to {\it three}
components to the dark matter density: axion dark matter produced via vacuum 
mis-alignment at the QCD phase transition, non-thermally produced axinos
from $\tz_1\to\ta\gamma$ decay (likely warm dark matter if $m_{\ta}\alt 1$ GeV),
and thermally produced axinos which are likely cold dark matter unless 
$m_{\ta}\alt 100$ keV. We compute the abundance of all three components of
dark matter in Yukawa-unified models for four different scenarios
containing either dominant axion or dominant axino cold dark matter. 
The relative abundances depend on $\Omega_{\tz_1}h^2$, $m_{\tz_1}$, 
$m_{\ta}$, $f_a/N$ and $T_R$. It is found that for all solutions with
$m_{16}\sim m_{3/2}\agt 5$ TeV, the value of $T_R$ needed is below limits
calculated from BBN constraints, thus solving the gravitino BBN problem.

We also find that it is very difficult for models
with PQ symmetry breaking scale $f_a$ lower than $\sim 2\times 10^{11}$ GeV
(case C1) to generate dominantly cold dark matter and a sufficiently large $T_R$
to be consistent with at least non-thermal leptogenesis, for any allowed value
of $m_{16}>5$ TeV. However, if the PQ scale is large enough ($\agt 4\times 10^{11}$ GeV), 
then we can generate a universe with axions as the dominant component of CDM (cases C2 and C3),
and only a smaller component is composed of possibly warm axinos with $m_{\ta}\alt 10^{-3}$
GeV. These solutions work out if $m_{16}\agt 10$ TeV, so that the higher side of the
$m_{16}\sim 3-15$ TeV range is preferred. This is also in accord with results presented
in Sec. \ref{ssec:bsg}, where it is found that $BF(b\to s\gamma )$ and 
$BF(B_s\to\mu^+\mu^- )$ decay also prefer $m_{16}\agt 10$ TeV. 
Finally, we show a case C4 with accidentally small axion CDM component, but which gives
cold thermally produced axino dark matter, as long as $m_{\ta}\agt 100$ keV. This scenario
allows for solutions with $m_{16}$ as low as 5  TeV.

As far as tests go of the Yukawa-unified SUSY models with mixed axion/axino dark matter,
it is clear from previous work that very large signal rates from gluino pair production and subsequent 
cascade decays should be visible at LHC relatively soon after start-up\cite{so10lhc}. The $\tz_1$
produced in cascade decay events will still yield events with $\eslt$, since the
$\tz_1\to \ta\gamma$ decay occurs far outside the detector. In cases C2 and C3, there is 
a good chance for direct detection of relic axion dark matter at experiments such as ADMX\cite{admx}.
However, direct and indirect searches for WIMP dark matter would likely turn up null results. 

\appendix

\section{Appendix: Yukawa-unified benchmark points from \Isajet\,7.79}
\label{app:bm}

Here, we present several updated benchmark points from \Isajet\,7.79
with Yukawa-unified solutions. Points A, B and C with
$m_{16}=5$, 10 and 15 TeV respectively all require mixed axion/axino
dark matter. Point H has $m_{16}\sim 3$ TeV, and can accomodate
neutralino dark matter since neutralino annihilation through the
$h$ resonance leads to a neutralino dark matter abundance in near accord with
WMAP5 measurements.\footnote{Here, we anticipate a theory error bar 
on the relic density calculation arising from a variety of
uncertainties to be of order 10-20\%.}

%
\begin{table}\centering
\begin{tabular}{lcccc}
\hline
parameter & Pt. A & Pt. B & Pt. C & Pt. H \\
\hline
$m_{16}$   & 5000   & 10000 & 15000 &  2859.53 \\
$m_{1/2}$  & 80.573 & 43.9442 & 15.2732 & 116.152 \\
$A_0$      & $-$10046.7 & $-$19947.3 & $-$32495.5 & $-$5582.68 \\
$m_{10}$   & 6273.26 & 12053.5 & 15098.8 & 3642.41 \\
$M_D$      & 1655.92 & 3287.12 & 5367 & 943.32 \\
$\tan\beta$& 49.3631 & 50.398 & 51.7337 & 48.281 \\
\hline
$f_t$      & 0.566 & 0.557 & 0.556 & 0.547 \\
$f_b$      & 0.560 & 0.557 & 0.549 & 0.500 \\
$f_\tau$   & 0.562 & 0.571 & 0.563 & 0.525 \\
$R$        & 1.01  & 1.02  & 1.03  & 1.09 \\
\hline
$\mu$      & 1101.2 & 3132.6 & 10365.8 & 440.4 \\
$m_{\tg}$   & 363.3 & 351.2 & 368.8 & 406.4 \\
$m_{\tu_L}$ & 4983.6 & 9972.1 & 14976.4 & 2855.7 \\
$m_{\tst_1}$& 833.7 & 2756.5 & 5422.5 & 318.6 \\
$m_{\tb_1}$ & 1322.0 & 3377.1 & 6020.3 & 802.1 \\
$m_{\te_L}$ & 4968.1 & 9940.6 & 14904.2 & 2841.0 \\
$m_{\twpm_1}$ & 109.2 & 116.4 & 136.2 & 115.7 \\
$m_{\tz_2}$ & 108.9 & 113.8 & 135.5 & 115.1 \\ 
$m_{\tz_1}$ & 49.8 &  49.2 & 54.4 & 56.6 \\ 
$m_A$       & 939.0 &  1825.9 & 4714.9 & 884.3 \\
$m_h$       & 124.1 &  127.8 & 128.4 & 115.0 \\ \hline
$\Delta a_\mu$ & $0.4\times 10^{-10}$ & $0.6\times 10^{-11}$ 
& $0.5\times 10^{-12}$ & $0.2\times 10^{-9}$\\
$BF(b\to s\gamma )$ & $1.9\times 10^{-4}$ & $3.0\times 10^{-4}$ 
& $3.1\times 10^{-4}$ & $1.3\times 10^{-4}$\\
$BF(B_s\to\mu^+\mu^- )$ & $2.9\times 10^{-8}$ & $8.1\times 10^{-9}$ & 
$4.3\times 10^{-9}$ & $3.6\times 10^{-8}$ \\
$\Omega h^2_{\tz_1}$ & 90.7 & 3881 & 522 & 0.13 \\
\hline
\end{tabular}
\caption{Masses and parameters in~GeV units
for four cases studies 
using \Isajet\,7.79 with $m_t=172.6$ GeV. 
We also list the $b\to s\gamma$ branching fraction, 
$\Delta a_\mu$ and $\Omega_{\tz_1}h^2$ for each case.
}
\label{tab:bm}
\end{table}
%

\acknowledgments

This research was supported in part by the U.S. Department of Energy
grant numbers DE-FG02-97ER41022.  
This work is also part of the French ANR project ToolsDMColl, BLAN07-2-194882.
SK thanks the Kavli Institute for Theoretical Physics China (KITPC) for 
hospitality during the final stage of this work. 
SS acknowledges financial support by Turkish Atomic Energy Authority.

%


\begin{thebibliography}{99}
%
\bibitem{gaugeunif} 
S. Dimopoulos, S. Raby and F. Wilczek, \prd{24}{1981}{1681};
U. Amaldi, W. de Boer and H. Furstenau, 
\plb{260}{1991}{447};
J. Ellis, S. Kelley and D. V. Nanopoulos, \plb{260}{1991}{131};
P. Langacker and Luo, \prd{44}{1991}{817}.
%
\bibitem{su5} H. Georgi and S. Glashow, \prl{32}{1974}{438}.
H. Georgi, H. Quinn and S. Weinberg, \prl{33}{1974}{451};
A. Buras, J. Ellis, M. K. Gaillard and D. V. Nanopoulos, \npb{135}{1978}{66}.
%
\bibitem{so10} H. Georgi, in {\it Proceedings of the American Institue   
of Physics}, edited by C. Carlson (1974); H. Fritzsch and P. Minkowski,  
Ann. Phys. {\bf 93}, 193 (1975); M. Gell-Mann, P. Ramond and R. Slansky,  
Rev. Mod. Phys. {\bf 50}, 721 (1978). For recent reviews,   
see R. Mohapatra, hep-ph/9911272 (1999) and S. Raby, in
Rept. Prog. Phys. {\bf 67} (2004) 755. For additional perspective, 
see G. Altarelli and F. Feruglio, \hepph{0405048}.
%
\bibitem{seesaw}  M. Gell-Mann, P. Ramond and R. Slansky, 
in {\it Supergravity, Proceedings of the Workshop}, Stony Brook, NY 1979 
(North-Holland, Amsterdam);  
T. Yanagida, KEK Report No. 79-18, 1979; 
R. Mohapatra and G. Senjanovic,  \prl{44}{1980}{912}.
%
\bibitem{old} B. Ananthanarayan, G.~Lazarides and Q.~Shafi, 
\prd{44}{1991}{1613} and \plb{300}{1993}{245}; 
G.~Anderson {\it et al.} \prd{47}{1993}{3702} and \prd{49}{1994}{3660};
V. Barger, M. Berger and P. Ohmann,   
\prd{49}{1994}{4908};
M. Carena, M. Olechowski, S. Pokorski and C. Wagner,  
Ref. \cite{hrs};   
B. Ananthanarayan, Q. Shafi and X. Wang, \prd{50}{1994}{5980};
R. Rattazzi and U. Sarid, \prd{53}{1996}{1553};
T.~Blazek, M.~Carena, S.~Raby and C.~Wagner, \prd{56}{1997}{6919}; 
T.~Blazek and S. Raby, \plb{392}{1997}{371};
T.~Blazek and S.~Raby, \prd{59}{1999}{095002};
T.~Blazek, S.~Raby and K.~Tobe, \prd{60}{1999}{113001}
and \prd{62}{2000}{055001}; 
H. Baer, M. Diaz, J. Ferrandis and X. Tata, 
\prd{61}{2000}{111701};
H. Baer, M. Brhlik, M. Diaz, J. Ferrandis,
P. Mercadante, P. Quintana and X. Tata, \prd{63}{2001}{015007};
S. Profumo, \prd{68}{2003}{015006}; C. Pallis, \npb{678}{2004}{398};
M. Gomez, G. Lazarides and C. Pallis, 
\prd{61}{2000}{123512}, \npb{638}{2002}{165} and \prd{67}{2003}{097701};
U. Chattopadhyay, A. Corsetti and P. Nath, \prd{66}{2002}{035003};
M. Gomez, T. Ibrahim, P. Nath and S. Skadhauge,
\prd{72}{2005}{095008};
K. Tobe and J. D. Wells, \npb{663}{2003}{123}.
%
\bibitem{exdimguts} Y. Kawamura, \ptp{105}{2001}{999}; 
G. Altarelli and F. Feruglio, \plb{511}{2001}{257}; 
L. Hall and Y. Nomura, \prd{64}{2001}{055003};  
A. Hebecker and J. March-Russell, \npb{613}{2001}{3};
A. Kobakhidze, \plb{514}{2001}{131}.
%
\bibitem{hrs} R.~Hempfling, \prd{49}{1994}{6168};  
L. J. Hall, R. Rattazzi and U. Sarid, \prd{50}{1994}{7048}; 
M.~Carena {\it et al.}, \npb{426}{1994}{269}.
%
\bibitem{bdr2} T. Blazek, R. Dermisek and S. Raby, \prd{65}{2002}{115004}.
%
\bibitem{bf} H. Baer and J. Ferrandis, \prl{87}{2001}{211803}.
%
\bibitem{abbbft} D. Auto, H. Baer, C. Balazs, A. Belyaev, J. Ferrandis 
and X. Tata, \jhep{0306}{2003}{023}.
%
\bibitem{bkss} H. Baer, S. Kraml, S. Sekmen and H. Summy, \jhep{0803}{2008}{056}.
%
\bibitem{isajet} F. Paige, S. Protopopescu, H. Baer and X. Tata, \hepph{0312045}; 
http://www.hep.fsu.edu/$\sim$isajet/
%
\bibitem{bdr1} T. Blazek, R. Dermisek and S. Raby, \prl{88}{2002}{111804}.
%
\bibitem{drrr} R. Dermisek, S. Raby, L. Roszkowski and
R. Ruiz de Austri, \jhep{0304}{2003}{037} and \jhep{0509}{2005}{029}.
%
\bibitem{bfpz} J. Feng, C. Kolda and N. Polonsky, \npb{546}{1999}{3}; 
J. Bagger, J. Feng and N. Polonsky, \npb{563}{1999}{3};
J. Bagger, J. Feng, N. Polonsky and R. Zhang, \plb{473}{2000}{264};
H. Baer,P. Mercadante and X. Tata, \plb{475}{2000}{289};
H. Baer, C. Balazs, M. Brhlik, P. Mercadante, X. Tata and Y. Wang,
\prd{64}{2001}{015002}; see also H. Baer, M. Diaz, P. Quintana and X. Tata, 
\jhep{0004}{2000}{016}.
%
\bibitem{auto} D. Auto, H. Baer and A. Belyaev and T. Krupovnickas, \jhep{0410}{2004}{066}.
%
\bibitem{wmap5} D.~N.~Spergel {\it et al.} (WMAP Collaboration), 
{\em Astrophys.~J.~Supp.}, {\bf 170} (2007) 377.
%
\bibitem{pq} R. Peccei and H. Quinn, \prl{38}{1977}{1440} and
\prd{16}{1977}{1791}.
%
\bibitem{ww} S. Weinberg, \prl{40}{1978}{223};
F. Wilczek, \prl{40}{1978}{279}.
%
\bibitem{axreview} For recent reviews on axion physics, see
J. E. Kim and G. Carosi, arXiv:0807.3125 (2008);
P. Sikivie, \hepph{0509198};
M. Turner, \prep{197}{1990}{67}.
%
\bibitem{nillesraby} H. P. Nilles and S. Raby, \npb{198}{1982}{102}.
%
\bibitem{axmass} E. J. Chun, J. E. Kim and H. P. Nilles, 
\plb{287}{1992}{123}.
%
\bibitem{ckkr} L. Covi, J. E. Kim and L. Roszkowski, \prl{82}{1999}{4180}; 
L. Covi, H. B. Kim, J. E. Kim and L. Roszkowski, \jhep{0105}{2001}{033}.
%
\bibitem{wilczek} K. Rajagopal, M. Turner and F. Wilczek, 
\npb{358}{1991}{447}.
%
\bibitem{steff_rev} For a recent review of axion/axino dark matter, see
F. Steffen, arXiv:0811.3347 (2008).
%
\bibitem{jlm} K. Jedamzik, M. LeMoine and G. Moultaka,
JCAP{\bf 0607} (2006) 010.
%
\bibitem{steffen} A. Brandenburg and F.~Steffen,
JCAP{\bf 0408} (2004) 008.
%
\bibitem{julien}
  A.~Boyarsky, J.~Lesgourgues, O.~Ruchayskiy and M.~Viel,
  arXiv:0812.0010 [astro-ph].
%
\bibitem{kohri} K. Kohri, T. Moroi and A. Yotsuyanagi, \prd{73}{2006}{123511};
for an update, see
M. Kawasaki, K. Kohri, T. Moroi and A. Yotsuyanagi, arXiv:0804.3745 (2008).
%
\bibitem{buchm} W. Buchmuller, P. Di Bari and M. Plumacher, 
Annal. Phys. {\bf 315} (2005) 305.
%
\bibitem{ntlepto} G. Lazarides and Q. Shafi, \plb{258}{1991}{305};
K. Kumekawa, T. Moroi and T. Yanagida, \ptp{92}{1994}{437};
T. Asaka, K. Hamaguchi, M. Kawasaki and T. Yanagida, \plb{464}{1999}{12}.
%
\bibitem{bs} H. Baer and H. Summy, \plb{666}{2008}{5}.
%
\bibitem{mtop} The Tevatron Electroweak Working group (CDF and D0 Collaborations), arXiv:0803.1683.
%
\bibitem{castano} D. Castano, Piard and P. Ramond, \prd{49}{1994}{4882};
A. Dedes, A. Lahanas and K. Tamvakis, \prd{53}{1996}{3793}.
%
\bibitem{box} A. Box and X. Tata, \prd{77}{2008}{055007} and arXiv:0810.5765 (2008).
%
\bibitem{bmpz} D. Pierce, J. Bagger, K. Matchev and R. Zhang, \npb{491}{1997}{3}.
%
\bibitem{mv} S. P. Martin and M. Vaughn, \prd{50}{1994}{2282}.
%
\bibitem{bfkp} H. Baer, J. Ferrandis, S. Kraml and W. Porod, \prd{73}{2006}{015010}.
%
\bibitem{kraml} B.C. Allanach, S. Kraml and W. Porod, \jhep{03}{2003}{016};
G. Belanger, S. Kraml and A. Pukhov, \prd{72}{2005}{015003};
S. Kraml and S. Sekmen in:  M.M. Nojiri {\it et al.},
{\it Physics at TeV Colliders 2007, BSM working group report}, 
arXiv:0802.3672 [hep-ph].
%
\bibitem{softsusy} B. Allanach, \cpc{143}{2002}{305}.
%
\bibitem{micromegas} G. Belanger, F. Boudjema, A. Pukhov and A. Semenov,
\cpc{149}{2002}{103}, \cpc{174}{2006}{577} and \cpc{176}{2007}{367}.
%
\bibitem{isatools} H. Baer, C. Balazs, A. Belyaev, J. K. Mizukoshi and X. Tata,
\jhep{0207}{2002}{050}.
%
\bibitem{bb} H. Baer and M. Brhlik, \prd{55}{1997}{3201}.
%
\bibitem{anlauf} H. Anlauf, \npb{430}{1994}{245}.
%
\bibitem{greub} C. Greub, T. Hurth and D. Wyler, \plb{380}{1996}{385}
and \prd{54}{1996}{3350}.
%
\bibitem{bbct} H. Baer, M. Brhlik, D. Castano and X. Tata, \prd{58}{1998}{015007}.
%
\bibitem{misiak} M.~Misiak {\it et al.}, \prl{98}{2007}{022002}.
%
\bibitem{bsg_ex} E.~Barberio {\it et al.} (Heavy Flavor Averaging Group), 
\hepex{0603003}.
%
\bibitem{alt} M. Albrecht, W. Altmannshofer, A. Buras, D. Guadagnoli and D. Straub,
\jhep{0710}{2007}{055}; W. Altmannshofer, D. Guadagnoli, S. Raby and
D. Straub, arXiv:0801.4363 (2008); D. Guadagnoli, arXiv:0810.0450 (2008).
%
\bibitem{tata} J. K. Mizukoshi, X. Tata and Y. Wang, \prd{66}{2002}{115003}.
%
\bibitem{cdf} T. Aaltonen {\it et al.} (CDF Collaboration),
\prl{100}{2008}{101802}.
%
\bibitem{absik} L. F. Abbott and P. Sikivie, \plb{120}{1983}{133};
J. Preskill, M. Wise and F. Wilczek, \plb{120}{1983}{127};
M. Dine and W. Fischler, \plb{120}{1983}{137};
M. Turner, \prd{33}{1986}{889}.
%
\bibitem{wss} H.~Baer and X.~Tata, {\it Weak Scale Supersymmetry: From 
Superfields to Scattering Events}, 
(Cambridge University Press, 2006).
%
\bibitem{ksvz} J. E. Kim, \prl{43}{1979}{103};
M. A. Shifman, A. Vainstein and V. I. Zakharov, \npb{166}{1980}{493}.
%
\bibitem{dfsz} M. Dine, W. Fischler and M. Srednicki, \plb{104}{1981}{199};
A. P. Zhitnitskii, \sjp{31}{1980}{260}.
%
\bibitem{braaten} E. Braaten and R. Pisarski, \npb{337}{1990}{569};
E. Braaten and T. C. Yuan, \prl{66}{1991}{2183}.
%
\bibitem{cremmer} E. Cremmer, S. Ferrara, L. Girardello and A. van Proeyen,
\npb{212}{1983}{413}.
%
\bibitem{nilles} H. P. Nilles, \prep{110}{1984}{1}.
%
\bibitem{relic_G} M. Bolz, A. Brandenburg and W. Buckmuller, \npb{606}{2001}{518};
J. Pradler and F. Steffen, \prd{75}{2007}{023509}.
%
\bibitem{bbn_gino} S. Weinberg, \prl{48}{1982}{1303};
R. H. Cyburt, J. Ellis, B. D. Fields and K. A. Olive, \prd{67}{2003}{103521};
K. Jedamzik, \prd{70}{2004}{063524};
M. Kawasaki, K. Kohri and T. Moroi, \plb{625}{2005}{7} and \prd{71}{2005}{083502}.
%
\bibitem{leptog} M. Fukugita and T. Yanagida, \plb{174}{1986}{45};
M. Luty, \prd{45}{1992}{455};
M. Luty, \prd{45}{1992}{455};
W. Buchm\"uller and M. Plumacher, \plb{389}{1996}{73} and \ijmpa{15}{2000}{5047};
R. Barbieri, P. Creminelli, A. Strumia and N. Tetradis, \npb{575}{2000}{61};
G. F. Giudice, A. Notari, M. Raidal, A. Riotto and A. Strumia, \hepph{0310123};
for a recent review, see W. Buchm\"uller, R. Peccei and T. Yanagida, 
\arnps{55}{2005}{311}.
%
\bibitem{NTlepto} G. Lazarides and Q. Shafi, \plb{258}{1991}{305};
K. Kumekawa, T. Moroi and T. Yanagida, \ptp{92}{1994}{437};
T. Asaka, K. Hamaguchi, M. Kawasaki and T. Yanagida, \plb{464}{1999}{12}.
%
\bibitem{epsilon} K. Hamaguchi, H. Murayama and T. Yanagida, \prd{65}{2002}{043512}.
%
\bibitem{bp} W. Buchm\"uller, P. Di Bari and M. Plumacher, \npb{643}{2002}{367}
and Erratum-ibid,{\bf B793} (2008) 362; \annp{315}{2005}{305} and \njp{6}{2004}{105}.
%
\bibitem{imy} M. Ibe, T. Moroi and T. Yanagida, \plb{620}{2005}{9}.
%
%
\bibitem{so10lhc} H. Baer, S. Kraml, S. Sekmen and H. Summy,
\jhep{0810}{2008}{079}.
%
\bibitem{admx} 
L. Duffy {\it et al.}, \prl{95}{2005}{091304} and \prd{74}{2006}{012006};
for a review, see S. Asztalos, L. Rosenberg, K. van Bibber, P. Sikivie
and K. Zioutas, \arnps{56}{2006}{293}.
%
\end{thebibliography}
\end{document}